\documentclass{aa}  
\usepackage{graphicx}
\usepackage{txfonts}
\usepackage[utf8]{inputenc}
\usepackage{advdate}
\usepackage[
colorlinks=true,
citecolor=blue,
]{hyperref}
\usepackage[switch,mathlines]{lineno}
\modulolinenumbers[2]

\newcommand{\mouno}[0]{\textit{Model~1}}
\newcommand{\modue}[0]{\textit{Model~2}}
\begin{document}

   \title{Dust distribution around low-mass planets on converging orbits}

   \subtitle{}

   \author{Francesco Marzari\inst{1}\fnmsep\thanks{Corresponding author.}
          \and
           Gennaro D'Angelo\inst{2}
          }

   \institute{Department of Physics and Astronomy, 
              University of Padova, via Marzolo 8, I-35131, 
              Padova, Italy\\
              \email{francesco.marzari@pd.infn.it}
         \and
              Theoretical Division, 
              Los Alamos National Laboratory, 
              Los Alamos, NM 87545, USA\\
              \email{gennaro@lanl.gov}
             }

   \date{Received ....; accepted ....}

 
  \abstract
  {Super-Earths can form at large orbital radii and migrate
  inward due to tidal interactions with the circumstellar disk. In this scenario,
  convergent migration may occur and lead to the formation of resonant pairs 
  of planets.}
  {We  explore the conditions under which convergent migration and 
  resonance capture take place, and what dynamical consequences can be expected 
  on the dust distribution surrounding the resonant pair. }
  {We combine hydrodynamic planet--disk interaction models with dust evolution
  calculations to investigate the signatures produced in the dust distribution 
  by a pair of planets in mean-motion resonances.  }
  {We find that convergent migration  takes place when the outer planet is the more
  massive. However,  convergent migration also depends  on the local properties of 
  the disk, and divergent migration may result as well. For similar disk 
  parameters, the capture in low degree resonances (e.g.,   2:1 or 3:2) 
  is preferred close to 
  the star where the resonance strength can more easily overcome the tidal 
  torques exerted by the gaseous disk.
  Farther away from the star, 
  convergent migration may result in  capture in high degree resonances. 
  The dust distribution shows potentially observable features typically when 
  the planets are trapped in a 2:1 resonance. 
  In other cases, with higher degree resonances (e.g., 5:4 or 6:5)
  dust features may not be sufficiently pronounced to be easily observable.}
  {The degree of  resonance established by a pair of super-Earths may be 
  indicative of the location in the disk where capture occurred. There can be significant differences in the dust distribution around a single super-Earth 
  and a pair of super-Earths in resonance.}

   \keywords{ Planet-disk interactions; Zodiacal dust}

   \maketitle
%
\section{Introduction}
In multi-planet systems close to their final formation stages, the convergent
migration of two planets with adjacent orbits may occur due to tidal interactions 
with the surrounding circumstellar gas. 
Convergent migration is an important aspect of the dynamical evolution in a 
multi-planet system since it can lead to resonant orbital configurations. 
If the resonant forcing generated by gravitational interactions between two
neighboring planets overcomes the tidal forcing, the resonant configuration 
can outlast the presence of the gaseous disk. Otherwise, the orbits can become 
so compact that they produce collision or scattering events \citep{marzaba2010,lega2013}. 
In the long term, dispersal of the disk may also destabilize a resonant
configuration.

In the case of two giant planets, the typical scenario for resonant capture is 
that of a more massive planet orbiting closer to the star than a less massive one.
For a pair of planets like  Jupiter and Saturn, for example, the exterior Saturn-mass body 
may not open a deep gap in the gas and its inward migration would be faster 
than that of the inner Jupiter-size planet, which is typically able to carve 
a deeper gap, and therefore drifts inward at a slower rate. Once the planets
are close enough to each other, they end up in a mean-motion resonance, 
which depends on the disk parameters
\citep{masset2001,lee2002,adams2005,thommes2005,beauge2006,crida2008,dangelo2012}.
The pair may then form and maintain a common gap, which can also significantly 
affect the surrounding distribution of dust \citep{marzari2019}. 

The migration of small planets, typically unable to open a gap in the gas, 
is solely affected by type~I torques, driven at corotation and Lindblad resonances 
\citep[e.g.,][and references therein]{tanaka2002,parde2010,parde2011}.
Inward convergent migration, in these cases, requires a more massive planet 
orbiting exterior to a less massive planet, since the type~I migration rate 
is linearly dependent on the mass of the body 
\citep[e.g.,][and references therein]{tanaka2002}.
In this study
we want to explore the evolution of two planets in the super-Earth to Uranus 
mass range in order to test whether convergent migration leads to resonance 
trapping, which mean-motion resonances are involved, and under what conditions 
they can occur.

\begin{figure}
\hspace{-0.5cm}
\includegraphics[width=\columnwidth]{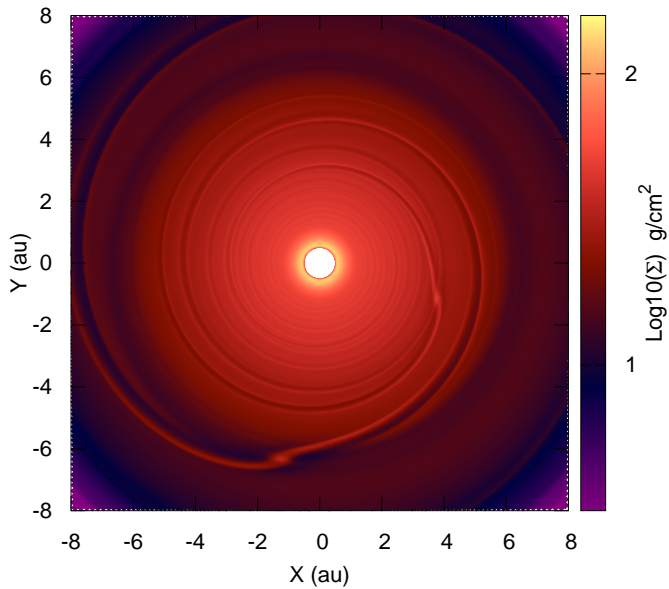}
\caption{\label{gas} Gas density distribution after $10$~Kyr 
from the beginning of a simulation with an initial surface 
density $\Sigma_0 = 50\,\mathrm{g\,cm^{-2}}$ at $1$~au. 
The position of the planets corresponds to the locations 
of the two gas ``overdense'' regions. 
}
\end{figure}

\cite{hands2014} explored the simultaneous migration of multiple planets 
forming at large orbital radii and migrating inward. They found that 
a large number of systems end up in resonance,  as expected according to
\cite{szuskie12}. They adopt an N-body model 
with artificial damping terms mimicking the torques due to tidal interactions 
with the gaseous disk. This approach, similar to that exploited by 
\citet{lee2002}, \citet{beauge2006}, and \citet{rein2012},
works reasonably well for massive 
planets in a type~II-like migration regime. It is also well suited 
for statistical studies of smaller planets because it relieves 
the computational overhead with respect to more sophisticated
methods.
However, in the case of a  small planet, it is not straightforward to assume 
that two planets embedded in a circumstellar disk migrate at the same 
rate as they would have in isolation. The perturbations induced by a second
body can have non-trivial effects. The wakes produced at 
the Lindblad resonances by one planet may interfere with those 
of the other planet, leading to a dynamical evolution for double planets that is different from that of  a single planet, 
possibly significantly different.
If the pair orbits in a compact configuration, even corotation torques 
may be affected by changes in the horseshoe dynamics of the gas in 
the proximity of each planet, due to the gravitational perturbations
of the other planet. 
As a consequence, the total torque on each planet may change depending 
on the orbits of the planets and on the characteristics of the disk. 
An example of wake superposition in the disk is shown in Figure~\ref{gas} 
and is discussed in more detail in the following sections. 

In order to perform a detailed exploration of convergent migration of
low-mass planets, we adopt here an approach based on hydrodynamics
simulations. This approach is more computationally expensive, but also
more accurate as it allows direct computations of the tidal torques
acting on the planets. 
A similar approach was used by \cite{szuskie05} while exploring 
the convergent migration of Earth-mass planets. Differently from what
is done in the present work, they considered high-mass disks, ranging from $0.5$ 
to $4$ times the mass of the ``minimum mass solar nebula'' with a piece-wise
gas surface density, a local isothermal equation of state, and inviscid gas. 
In this paper we use an energy equation for the gas that accounts for heating
and cooling and for lower gas densities, as are expected toward the final stages
of formation. We also use a viscous gas, and apply different prescriptions 
for the viscosity.
Contrary to \cite{szuskie05}, we find that convergent migration is not 
always attained when the outer planet is more massive than the inner one, 
and that the type of resonance in which the planets are locked generally
depends on the disk properties and planet location.
Furthermore, this computational method also allows us to compute 
the evolution of dust particles populating the protoplanetary disk and 
to search for signatures that the two planets may leave in the dust 
distribution as a result of their resonant configuration. 
This  modeling effort is particularly relevant in light of the complex 
morphology of circumstellar disks observed by ALMA.

In Sect.~\ref{sec:numerical} we describe the numerical algorithm adopted
here to model the evolution of the disk, planets, and dust grains. 
The dynamics of a pair of planets that approach each other close 
to the star and are trapped in low degree resonances (e.g., 2:1 and
3:2) is discussed in Sect.~\ref{sec:model1}. 
In Sect.~\ref{sec:model2} we model planets starting resonances, and encountering 
them, farther away from the star, and in which we find 
trapping in higher-degree resonances, such as  5:4 and 6:5. 
We also show cases of divergent migration with similar initial 
conditions, but occurring at high disk densities. 
Section~\ref{sec:dust} is dedicated to the exploration of the dust features 
in the proximity of two resonant planets. Finally, in Sect.~\ref{sec:discussion} 
we summarize and discuss the implications of our results. 

\section{Numerical model}
\label{sec:numerical}

The evolution of the gaseous disk, of the two planets, and of dust 
particles is computed with the 2D FARGO code \citep{Masset}, 
as modified by \cite{picogna2018} to include the particle dynamics. 
We performed simulations in which the energy equation contains
viscous heating and radiative cooling through the disk surface:
\begin{equation}
\frac{\partial E}{\partial t} + \nabla \cdot (E\,\mathbf{u})
                =  -P \, \nabla\cdot\mathbf{u} + Q^{+} - Q^{-}.
\label{eq:eneq}
\end{equation}
Here $E$ and $P$ are the total energy (surface) density and 
pressure, respectively, and $\mathbf{u}$ is the gas velocity field.
The quantity $Q^{+}$ is the viscous dissipation term computed from 
the components of the viscosity stress tensor \citep[]{m&m,dangelo2003}.
The term 
$Q^{-} = 2 \sigma_{\mathrm{SB}} T_{\mathrm{eff}}^4$ 
represents the local radiative cooling ($\sigma_{\mathrm{SB}}$ is
the Stefan--Boltzmann constant) in which the effective
temperature $T_{\mathrm{eff}}$ depends on the vertical optical 
thickness of the disk and is computed by exploiting the 
Rosseland mean opacity $\kappa$ \citep{linbell}, as in 
\citet[]{dangelo2003}.

In the simulations we model the evolution of $4\times 10^{5}$ 
dust particles with radii 
$10 \,\mathrm{\mu m}$, 
$100 \,\mathrm{\mu m}$, 
$1 \,\mathrm{mm}$, and 
$1 \,\mathrm{cm}$. 
Their trajectories are integrated taking into account the gas and 
the planetary perturbations.
We adopted these sizes since they are 
important in determining peculiar features of circumstellar disks
that can be potentially detected, for example by  ALMA in the (sub)mm band. 

The aerodynamic forces acting on the dust particles are computed 
as in \cite{picogna2015}. We briefly summarize the main 
forces determining the dynamical behavior  of the dust grains. 
The drag force acting on a spherical dust particle of radius $s$, 
moving with a velocity $\mathbf{v}$ relative to the gas 
is written as 
\begin{equation}
\mathbf{F}_D = (1-f)\,\mathbf{F}_{D,\mathrm{E}} + f\, \mathbf{F}_{D,\mathrm{S}}
\label{eq:FD}
,\end{equation} 
where 
\begin{equation}
\mathbf{F}_{D,\mathrm{E}}  =  -\frac{4}{3}\pi s^{2} \rho_{g} v_\mathrm{th} \mathbf{v} \label{eq:FDE}
\end{equation}
\vskip -0.4 truecm
\begin{equation}
\mathbf{F}_{D,\mathrm{S}}  =  -\frac{1}{2}\pi s^{2} C_D \rho_{g} v \mathbf{v}
\label{eq:FDS}
\end{equation}
are the drag forces in the Epstein and Stokes regimes, respectively. 
In Equation~(\ref{eq:FDE})
$v_\mathrm{th} = \sqrt{8 k_{\mathrm{B}} T/(\pi \mu m_\mathrm{H}}) $ is 
the mean thermal velocity of the gas molecules, $T$ the local temperature 
of the gas, $\rho_{g}$ the gas volume density, $m_\mathrm{H}$ the hydrogen 
atom mass, and $\mu$ the mean molecular weight of the gas. 

The drag force in the Stokes 
regime is proportional to the drag coefficient $C_D$, whose value is
taken from \citet{weidenschilling1977} and depends on the Reynolds number.
The transition between the two drag forces is determined by the coefficient 
$f$, given by 
\begin{equation}
f = {s \over {s + \lambda} } = {1 \over {1 + \mathrm{Kn}}},
\end{equation}
where $\lambda$ is the mean free path of the gas molecules and 
$\mathrm{Kn} = \lambda/s$ is the Knudsen number. Comparing the 
expressions for $\mathbf{F}_{D,\mathrm{E}}$ and $\mathbf{F}_{D,\mathrm{S}}$, 
it can be shown that they are equal when $\mathrm{Kn} = 4/9$ for Reynolds 
numbers $<1$ \citep[see][]{weidenschilling1977}. 

Due to drag forces, particles experience a radial drift relative to 
the gas;  in particular, they  respond to density and velocity 
gradients in the gas. The drift velocity (relative to the gas), 
in conditions of a stationary gas surface density, can be approximated 
as \citep{birnstiel2010,pinilla2012}
\begin{equation}
v_\mathrm{drift} = \frac{1}{\mathrm{St}^{-1} + \mathrm{St}}%
                   \left(\frac{\partial P}{\partial r}\right)%
                   \frac{1}{\Omega \rho_g},
\end{equation}
where $\Omega=\Omega(r)$ is the disk's Keplerian frequency.
Indicating with $m_{s}$ the mass of a particle,
$\mathrm{St}=m_{s} v \Omega/\mathbf{F}_{D}$ represents the Stokes number 
(sometimes referred to as non-dimensional stopping time) and $\Omega$ 
the Keplerian frequency of the disk. Since $v_\mathrm{drift}$ depends 
on the radial derivative of the gas pressure $P$, any local pressure 
maximum in the disk (with a significant azimuthal extent) will collect 
and trap grains, both orbiting in the vicinity of the maximum and those 
drifting inward from farther disk regions. 

The dust particles are initially distributed with a surface density 
that is constant in azimuth around the star and declines as the inverse
of the orbital distance, $\sigma(t=0)\propto 1/r$. This density distribution 
can be re-normalized to a different value at a reference radius (e.g., at 
$1\,\mathrm{au}$) since the particles do not interact among themselves
and there is no back-reaction of the dust on the gas. Therefore, the particles 
can  be thought of as tracers of the dynamical evolution of the dust population.  The particles that cross the inner border of the disk grid are re-initialized 
at the outer border.

\subsection{Disk initial setup}

\begin{figure*}
\resizebox{\linewidth}{!}{\includegraphics[]{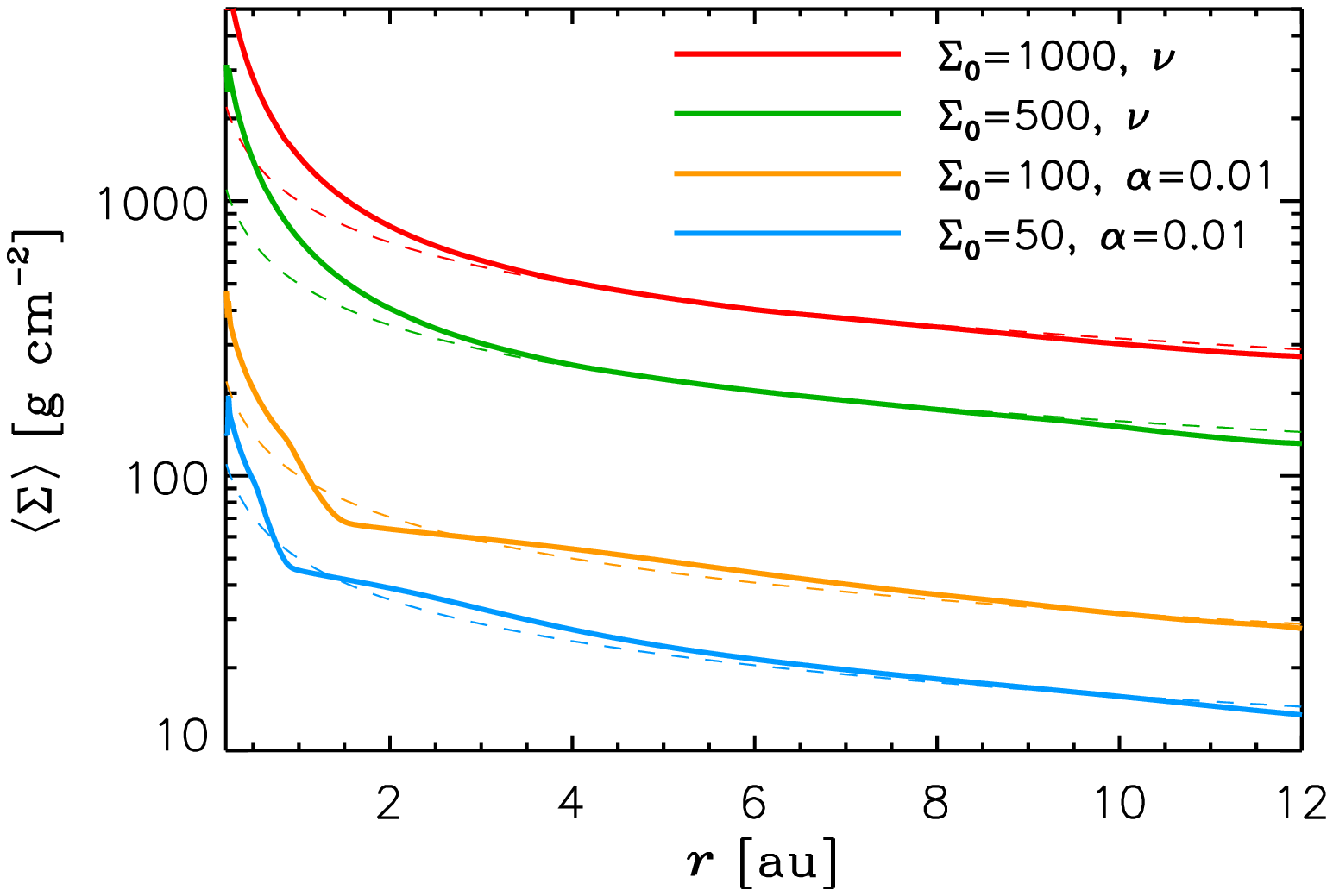}%
                          \includegraphics[]{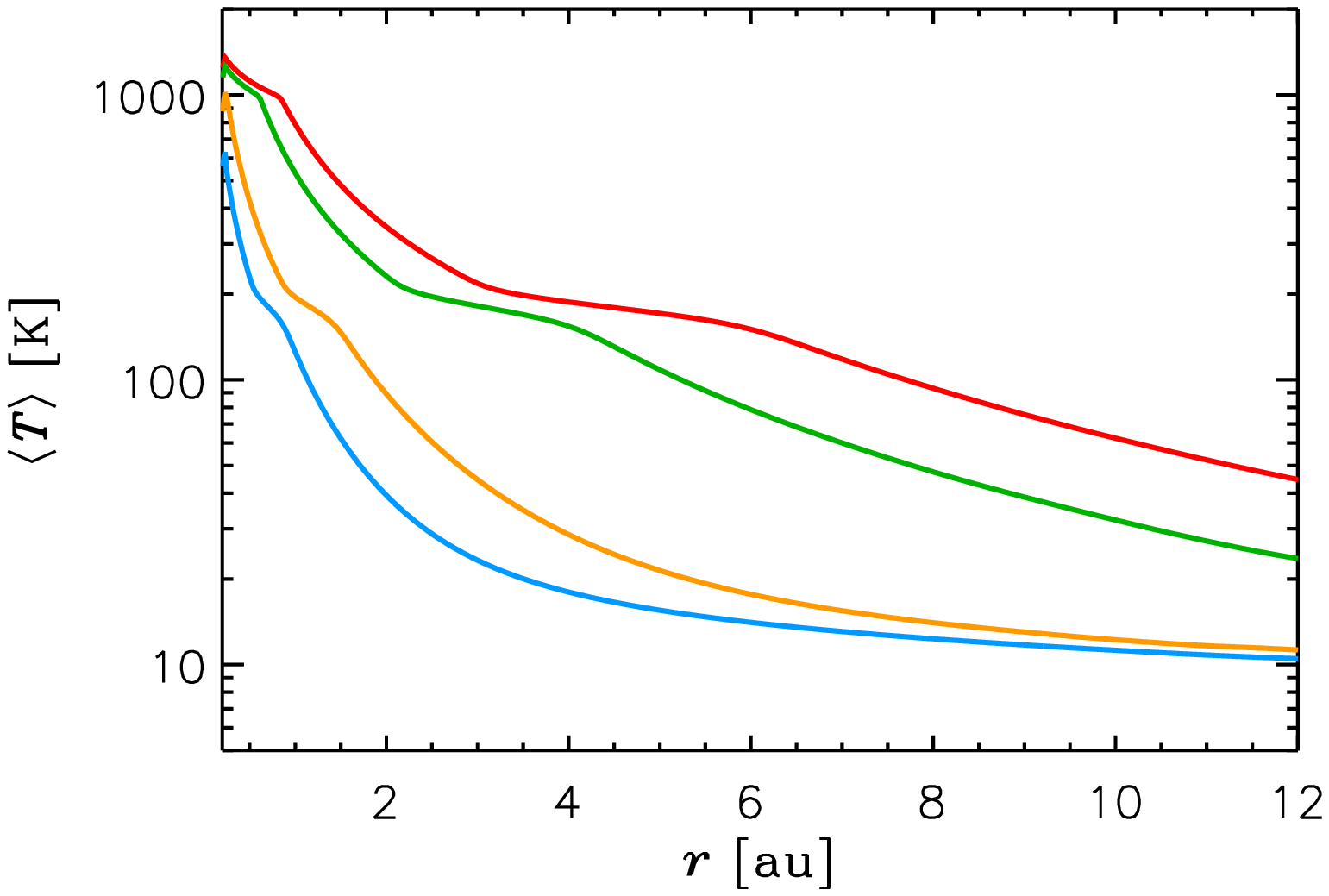}}
\caption{\label{fig:nop}
Evolution of the radial profiles of the disks. 
Left: Average gas surface density of the unperturbed disks after 
$12$~$\mathrm{kyr}$ (solid lines) compared to the initial 
distributions (thinner dashed lines) for various model setups, 
as indicated in the legend. The higher density and warmer models 
have constant kinematic viscosity $\nu$ (see text).
Right: Average gas temperature for the same models as in the
left panel.}
\end{figure*}

In modeling the circumstellar disk with FARGO, we
adopt a polar grid $(r,\phi),$   typically with $682\times 512$ elements 
uniformly covering the disk, which extends from $r = 0.5\,\mathrm{au}$ 
to $12\,\mathrm{au}$ from the star and $2\pi$ in azimuth. 
The initial surface gas density scales as 
\begin{equation}
\Sigma (r) = \Sigma_0 \left(\frac{r}{\mathrm{au}}\right)^{-p},
\label{eq:Sigma_r}
\end{equation}
with the power law index fixed to $p=0.5$. We adopt different values 
for $\Sigma_0$, smaller than those predicted for the minimum mass 
solar nebula \citep{weidenschilling1977,hayashi81}. 
We focus our study on the migration of super-Earths under the 
hypothesis that they formed at orbital radii of a few astronomical 
units or beyond, and are close to their final mass. The disk is
therefore expected to be old and to have partially dissipated 
via viscous evolution and photo-evaporation (and winds in general). 
We consider values of $\Sigma_{0}$ ranging from $50$ to
$1000\,\mathrm{g\,cm^{-2}}$. 

The disk aspect ratio is initialized to $h = H/r = 0.036$, and 
two different prescriptions for the kinematic viscosity $\nu$
are applied. 
We adopt either a constant $\alpha$-viscosity parameter, with values
ranging from $0.001$ to $0.01$, or a constant value 
$\nu = 10^{-5}$~$r_{0}^{2}\Omega_{0}$ ($\Omega_{0}$ is $\Omega$ at
$r_{0}=1$~$\mathrm{au}$). 
These two viscosity prescriptions can produce non-trivial
differences in the evolution of the gaseous disk. The former prescription leads to a kinematic viscosity
$\nu\propto h^{2}\sqrt{r}$, an increasing function of 
the radius if $h$ is constant in radius or the disk is flared.
The latter prescription corresponds instead to a variable $\alpha$
situation, with $\alpha\sim 0.01$ at $1$~$\mathrm{au}$ for
$h$ in the range $0.02$-$0.04$.

Figure~\ref{fig:nop} shows the unperturbed surface density
(averaged in azimuth around the star)
of some disk models on the left, after they settle in a steady state 
(solid curves), along with the initial profiles (dashed curves).
The corresponding temperatures are shown on the right.
The results indicate that there is only marginal evolution of $\Sigma$
inside a few $\mathrm{au}$, where the density profile tends to steepen 
somewhat. 
Otherwise, the surface density power index remains nearly constant at 
$p\approx 0.5$ during the evolution.
Other unperturbed models were also calculated and behave similarly.

The temperature profile of the disk is an important quantity in type~I 
migration since it affects the torque experienced by the planet 
\citep[see, e.g.,][and references therein]{kley2012}.
Neglecting the planetary perturbations, when the disk reaches a near-steady 
state and is vertically optically thick, heating and cooling are in 
equilibrium at a temperature such that
\begin{equation}
T^{2}\approx \left(\frac{27}{128}\frac{\kappa\nu}{\sigma_{\mathrm{SB}}}\right)^{1/2} \Sigma\Omega.
\label{eq:T2_thik}
\end{equation}
In the range of temperature between $\approx 200$~$\mathrm{K}$ and 
several hundred $\mathrm{K}$, \citeauthor{linbell}'s opacity does not 
vary much. Therefore, in the models with constant $\nu$ we expect
$T\propto 1/r$ (where $p\approx 0.5$). 
At $T\approx 100$~$\mathrm{K}$ and below, $\kappa\propto T^{2}$ 
and the gas temperature is expected to be approximately proportional 
to $1/r^{2}$.
We note that $\kappa\propto T^{2}$ at low temperatures
when micron-sized ice grains dominate the
opacity (see \citeauthor{pollack1985} \citeyear{pollack1985}.)
If the disk is vertically optically thin (e.g., $\kappa\Sigma\ll 1$), 
then at equilibrium
\begin{equation}
T^{2}\approx \left(\frac{9}{16}\frac{\nu}{\sigma_{\mathrm{SB}}\kappa}\right)^{1/2} \Omega.
\label{eq:T2_thin}
\end{equation}
In the $\alpha$-viscosity models, the kinematic viscosity is 
$\nu\propto \alpha T/\Omega$, hence $T\propto r^{-3/10}$ for 
$T\ll 100$~$\mathrm{K}$. 
These approximate scaling behaviors are in agreement with the curves 
in the right panel of Figure~\ref{fig:nop}, in the appropriate 
temperature ranges.

For practical purposes, quasi-steady states of density and 
of temperature are attained at much earlier times than those 
displayed in Figure~\ref{fig:nop}.
In all simulations that include planets, we run the numerical 
models for $300$ orbital periods of the inner planet, by keeping
the planets' orbits fixed, allowing the system to relax. 
We then release both planets, whose orbits can evolve from 
then onward under the influence of the disk's gravitational
torques. In the plots presented here, only the free evolution 
of the planets (i.e., from the time of release) is shown. 

\subsection{Initial planetary configurations}

We consider two different initial configurations for the planets. 
In the first, \mouno, the inner planet, whose mass is 
$m_1 = 5\,M_{\oplus}$, initially moves on a circular orbit at 
$a_1 = 2$~au, while the outer planet of mass $m_2 = 15\,M_{\oplus}$, 
is placed on a circular orbit at $a_2 = 3.3$~au. 
In the second configuration, \modue, the mass of the planets is 
the same as in \mouno,\ but we start with 
$m_1$ at $a_1 = 4$~au and $m_2$ at $a_2 = 6.5$~au. In both setups 
the two planets begin their orbital evolution outside of the 2:1 
mean-motion resonance.

In the two models we expect a different balance between the resonance 
strength and the torque exerted by the gaseous disk. 
Following the strength criterion adopted by \citet{murray-dermottSS} 
to describe the power relation between spin-orbit resonances and 
tidal torques on a satellite, we can compare the scaling of the 
mean motion resonance (hereinafter MMR) 
strength with that of the disk torque on the planets.
The former, according to \citet{murray-dermottSS}, in a simplified 
three-body problem scales as the mean motion squared (i.e., as $r^{-3}$). 
Consequently, while approaching the star the resonance forcing  acting
on the planets increases. The torque term due to the gaseous disk scales
as
\begin{equation}
\Gamma_{0} = \frac {q}{h^2} \Sigma r^4 \Omega^{2},
\label{eq:mig_vel}
\end{equation}
where $q$ is the planet-to-star mass ratio. If the disk's aspect ratio 
is nearly independent of (or weakly dependent on) the radial distance, 
the torque $\Gamma_0$ scales as $r^{1/2}$ (since $\Sigma\propto r^{-1/2}$) 
and it increases outward. As a consequence, the balance between the 
resonance strength and the tidal torque sways in favor of the former 
(and hence of low degree MMR locking of the orbits) as the planets move 
closer to the star. Higher degree (i.e., more compact) resonant 
configurations are expected at larger orbital distances.
We note that  the same outcome would be expected if $\Gamma_0$ were
nearly independent of $r$, for example if the disk had a typical flaring
index $d\ln{h}/d\ln{r}\approx 2/7$ \citep[e.g.,][]{chiang2010}.

\section{\mouno: low degree resonances}
\label{sec:model1}

\begin{figure}
\includegraphics[width=\columnwidth]{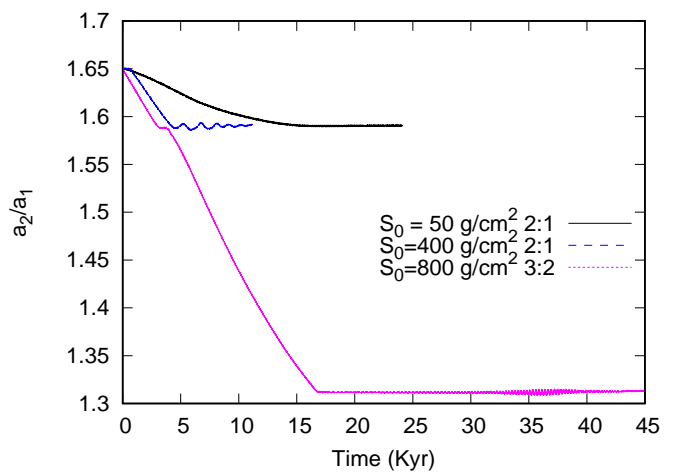}
\caption{\label{2au8} 
Capture in the 2:1 (black and blue solid line) and 3:2 (magenta solid line) 
MMRs of the two planets resulting from converging migration in \mouno. 
At the resonance, the ratio $a_2/a_1$ becomes constant.  
The initial surface density of the gas is $\Sigma_0 = 50$
and $400\,\mathrm{g\,cm^{-2}}$ for the capture in the 2:1 MMR and 
$\Sigma_0 = 800\,\mathrm{g\,cm^{-2}}$ for the capture in the 3:2 MMR
(see legend). 
The turbulence viscosity parameter in these cases is $\alpha = 0.01$. 
}
\end{figure}

In \mouno, for low values of the surface densities, $\Sigma_0 = 50$ 
and $400\,\mathrm{g\,cm^{-2}}$, and for a turbulence viscosity 
corresponding to $\alpha = 0.01$, the two planets become trapped in 
the 2:1 MMR. 
The torque exerted on the planets by the gaseous disk is not strong 
enough to overcome the  resonant forcing once the planets reach this 
resonance, and their orbits are therefore locked. 
For a higher density, $\Sigma_0 = 800\,\mathrm{g\,cm^{-2}}$, 
the planets cross the 2:1 MMR and are permanently captured in 
the 3:2 MMR. After capture, the planets migrate inward at the 
same speed. This is illustrated in Figure~\ref{2au8}, where 
we show the evolution in time of the ratio between the semi-major 
axis of the outer planet $a_2$ and that of the inner planet $a_1$.

\begin{figure}
\includegraphics[width=\columnwidth]{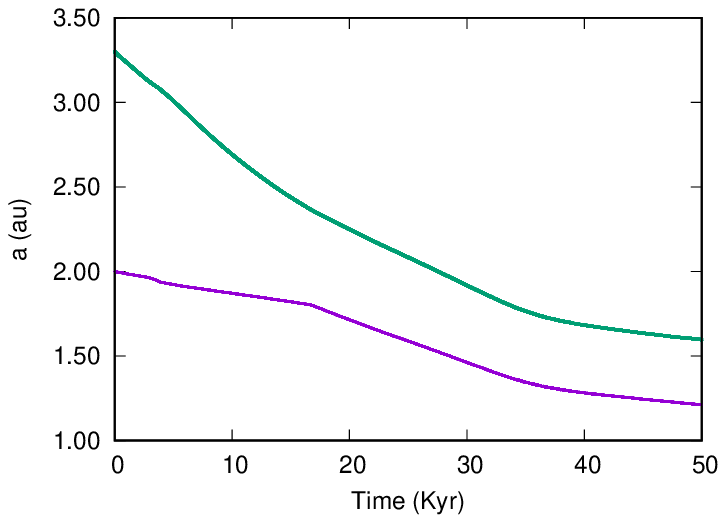}
\includegraphics[width=\columnwidth]{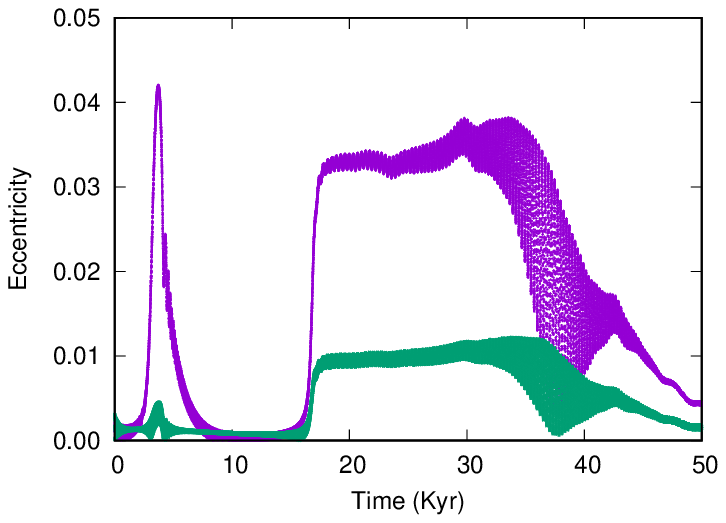}
\includegraphics[width=\columnwidth]{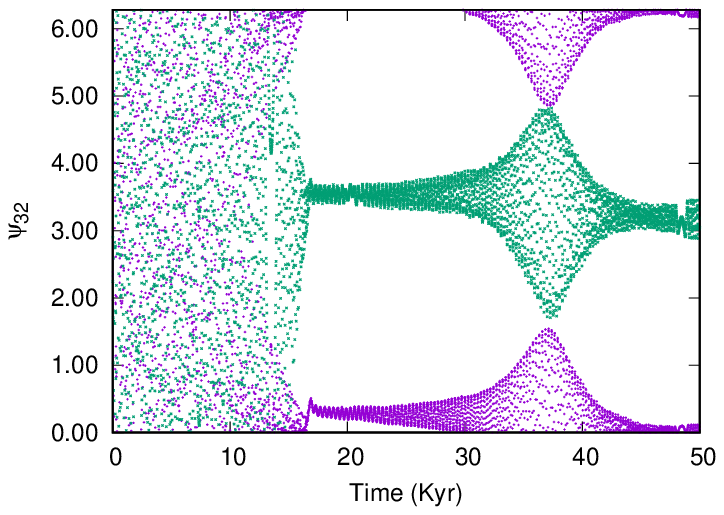}
\caption{\label{2au32res} 
Capture and evolution of two planets in  3:2 resonance. The panels show 
the time evolution of the semi-major axis (top), the orbital eccentricity 
(middle), and the critical arguments of the resonance  
$\psi^{i}_{32}$ in Equations~(\ref{eq:psi321}) and (\ref{eq:psi322}) 
(bottom).  
}
\end{figure}

A noteworthy behavior is that of the case ending with the capture in 
the 3:2 MMR. In Figure~\ref{2au32res} we display the evolution of 
the semi-major axis, eccentricity, and critical arguments of the system. 
As expected, the outer and more massive planet migrates faster than 
the inner one until it approaches the resonance location. 
During the evolution, the pair is temporarily trapped in 
 2:1 resonance (see Figure~\ref{2au8}), an event marked by 
a large jump in the orbital eccentricity of both planets 
(and especially of the inner one). 
The 2:1 MMR is eventually crossed. Once the planets become trapped 
in the 3:2 MMR, they begin to migrate at the same speed, which is 
intermediate between their migration velocities prior to resonance 
capture. While in resonance the orbital eccentricity of both 
planets rapidly grows and remains high for an extended period of time 
during which the two critical arguments of the resonance librate: 
\begin{eqnarray}
\psi^{1}_{32} & = & 3 \lambda_2 - 2 \lambda_1 - \varpi_1 \label{eq:psi321}\\
\psi^{2}_{32} & = & 3 \lambda_2 - 2 \lambda_1 - \varpi_2 \label{eq:psi322}.
\end{eqnarray}
We note that $\lambda_i$ is the true longitude of the planet, 
whereas $\varpi_i$ is the longitude of pericenter, for $i=1$ 
(the inner planet)  and $2$ (the outer planet).
The critical angles are slightly shifted with respect to 
the predicted values of $0$ and $\pi$ because of dissipative 
effects driven by the gas, but they show the symmetric apsidal 
corotation predicted by \cite{beauge2006}.  
When the inner planet approaches $\approx 1.3$~au, 
there is a change in the dynamical evolution of the pair, 
possibly due to the increase in the surface density gradient 
of the gas distribution (see Figure~\ref{fig:nop}). 
The migration speed slows down, as expected in a disk with slope 
$p\approx 1/2$, and the eccentricity is damped to small values. 

\section{\modue: high degree resonances}
\label{sec:model2}

\begin{figure}
\includegraphics[width=\columnwidth]{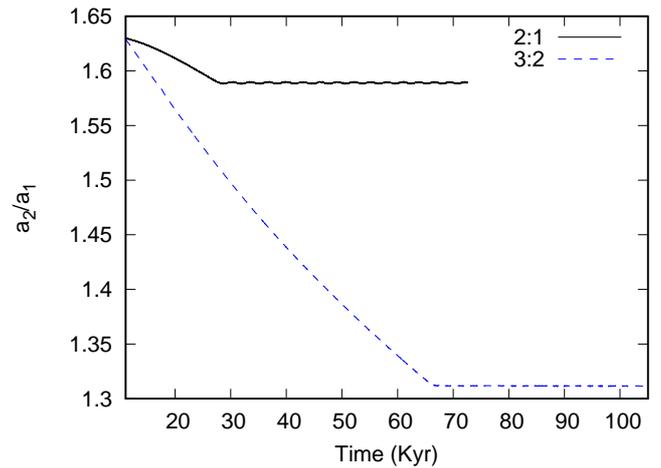}
\caption{\label{lowdens} 
Capture in the 2:1 (black solid line) and 3:2 (green dashed line) MMRs of 
the two planets, resulting from converging migration in \modue. 
The initial surface density of the gas is $\Sigma_0 = 50\,\mathrm{g\,cm^{-2}}$ 
for the capture in the 2:1 MMR and $\Sigma_0 = 100\,\mathrm{g\,cm^{-2}}$ for 
the capture in the 3:2 MMR. The turbulence viscosity parameter in these cases
is $\alpha = 0.01$. 
}
\end{figure}

In \modue, where the planets begin to migrate when they are farther 
away from the star, the behavior is significantly different. 
As anticipated above, for similar disk conditions the pair is expected 
(in general, but not always) to establish more compact orbital configurations
prior to capture.
When a viscosity $\alpha = 0.01$ is applied for the low density 
value $\Sigma_0 =  50\,\mathrm{g\,cm^{-2}}$, we find again capture in 
the 2:1 MMR.  Already at a density as low as $\Sigma_0 =  100\,\mathrm{g\,cm^{-2}}$, the gravitational torques exerted by 
the gas can drive the planets across the 2:1 MMR, and they become trapped in 
 3:2 resonance (see Figure~\ref{lowdens}). 
In \mouno\ this outcome is found for a significantly higher density, 
$\Sigma_0 =  800\,\mathrm{g\,cm^{-2}}$ (see Figure~\ref{2au8}), while for 
$\Sigma_0 =  400\,\mathrm{g\,cm^{-2}}$ capture in the 2:1 MMR is still 
the likely outcome. 

\begin{figure}
\includegraphics[width=\columnwidth]{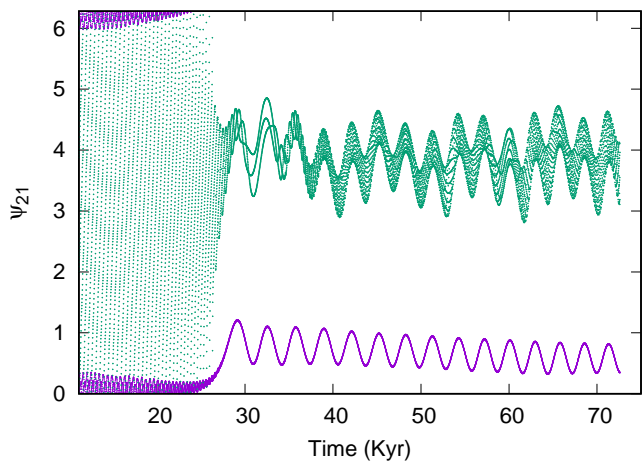}
\caption{\label{res21} 
Critical arguments of the 2:1 MMR as a function of time 
($\psi^{1}_{21} = 2 \lambda_2 - 1 \lambda_1 - \varpi_1$ and 
$\psi^{2}_{21} = 2 \lambda_2 - 1 \lambda_1 - \varpi_2$). 
The libration centers are shifted with respect to 0 and $\pi$ due to 
the effects of gas perturbations.
}
\end{figure}

As argued in Sect.~\ref{sec:numerical}, the different dynamical behavior 
of \modue\ with respect to \mouno\ is due to the different balance between
the tidal torques acting on the planets and the resonant forcing.  
The strength of the gas perturbations is expected to be greater in the 
orbital configuration realized by \modue\ (based on the torque strength
corresponding to the applied $\Sigma$), and this is further confirmed 
by the larger offset of the resonance libration centers with respect to 
$0$ and $\pi$. This is shown in Figure~\ref{res21}, where the two critical 
arguments of the resonance are shown. At the beginning of the simulation it  appears 
that $\psi^{1}_{21} = 2 \lambda_2 - 1 \lambda_1 - \varpi_1$ is already 
librating around $0$, but this is 
only due to the sampling interval as the output interval of
the calculation is not short enough. In reality, the critical argument
circulates with $\psi^{1}_{21}$ varying rapidly between $0$ and $\pi$. 
This behavior is not caught in the plot.

\begin{figure}
\includegraphics[width=\columnwidth]{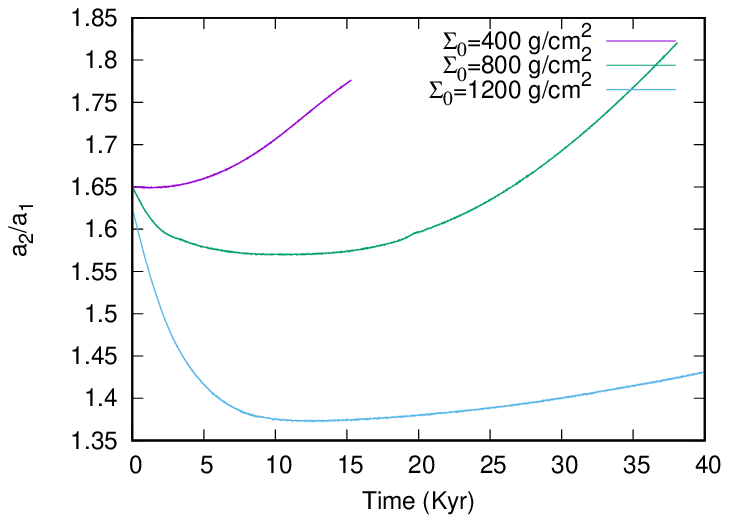}
\includegraphics[width=\columnwidth]{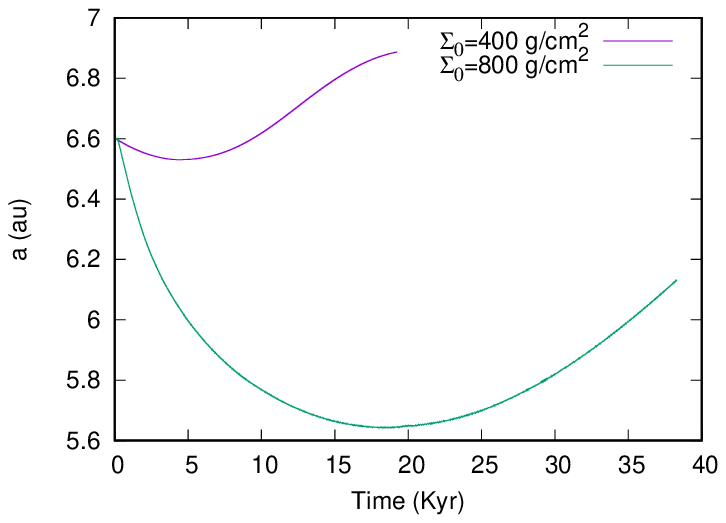}
\includegraphics[width=\columnwidth]{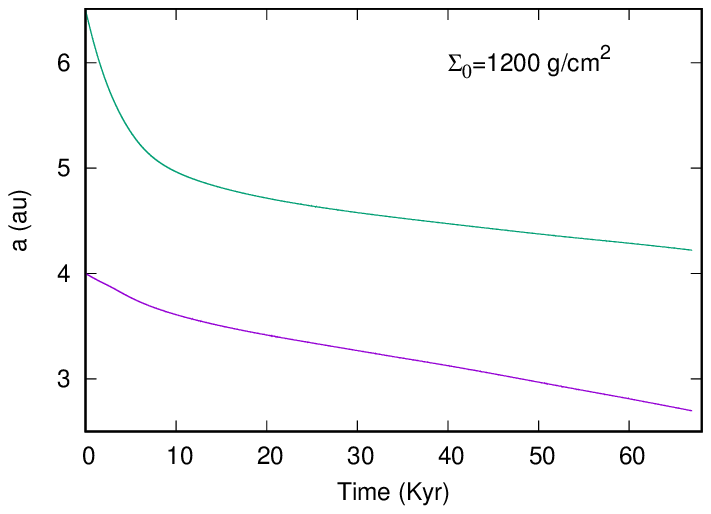}
\caption{\label{divergent} Divergent migration of two planets 
resulting from surface density values 
$\Sigma_0 = 400, 800, 1200\,\mathrm{g\,cm^{-2}}$ and viscosity 
$\alpha = 0.01$ (top panel). In the middle panel the evolution 
of the outer planet is shown for the cases with $\Sigma_0 = 400$ and 
$800\,\mathrm{g\,cm^{-2}}$, as indicated. In both cases, after 
an initial phase of inward migration, the planet drifts outward
(but the longer-term behavior was not investigated). 
In the bottom panel the divergent migration is instead due to 
the outer planet slowing down. In all cases, the inner planet keeps
migrating inward.
}
\end{figure}

When the surface density of the disk is progressively increased above 
$\Sigma_0 = 100\,\mathrm{g\,cm^{-2}}$ for the same value of viscosity 
($\alpha = 0.01$), convergent migration no longer occurs. 
As shown in Figure~\ref{divergent}, when the density is 
$\Sigma_0 = 400\,\mathrm{g\,cm^{-2}}$ and higher, the planets' migration
paths diverge. This may be due either to the onset of outward 
migration of the outer planet (middle panel of Figure~\ref{divergent}) or to 
the outer planet slowing down its inward migration (bottom panel of 
Figure~\ref{divergent}).
This outcome may occur for various reasons related to the balance between 
the Lindblad and corotation torques, which may also be altered by 
density perturbations due to gravitational interactions between 
the two planets. We do not investigate the physical and orbital conditions 
under which divergent migration may occur, but we note that (under 
the conditions investigated here) convergent migration is not guaranteed 
even if the outer planet is more massive than the inner one.

\begin{figure}
\includegraphics[width=\columnwidth]{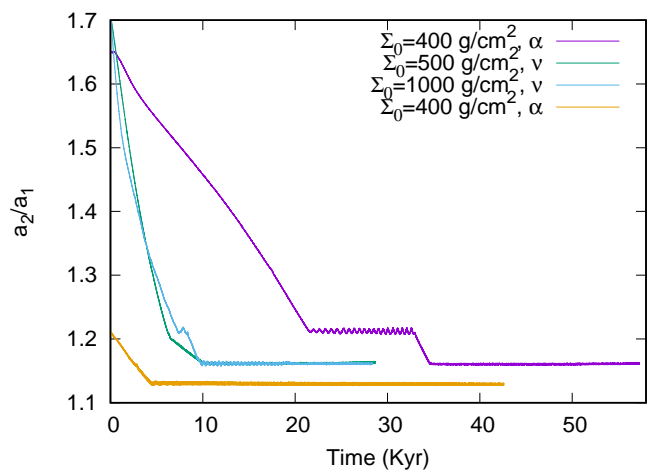}
\caption{\label{highdens} 
Capture in  5:4 resonance in three different 
configurations. The first has $\Sigma_0 = 400\,\mathrm{g\,cm^{-2}}$ 
and viscosity corresponding to $\alpha = 0.001$, the second 
$\Sigma_0 = 500\,\mathrm{g\,cm^{-2}}$ and constant kinematic viscosity 
$\nu = 10^{-5}$, and the third $\Sigma_0 = 1000\,\mathrm{g\,cm^{-2}}$ 
and $\nu = 10^{-5}$. The bottom curve shows capture in  6:5 resonance 
with $\Sigma_0 = 400\,\mathrm{g\,cm^{-2}}$ and viscosity equal to 
$\alpha = 0.001$. 
In this last case, however, the planets start on close orbits, just
exterior to the 4:3 MMR.
}
\end{figure}

\begin{figure}
\includegraphics[width=\columnwidth]{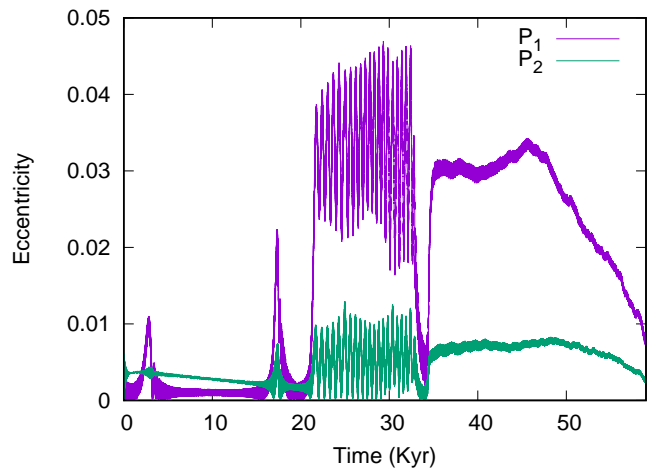}
\caption{\label{ecce54} Eccentricity evolution of the two planets 
in the case with $\Sigma_0 = 400\,\mathrm{g\,cm^{-2}}$ and viscosity 
$\alpha = 0.001$. Before capture each resonance crossing is 
characterized by a large jump in orbital eccentricity.
}
\end{figure}

To restore conditions conducive to convergent migration and resonance 
capture it was necessary  either to reduce the viscosity parameter $\alpha$ to 
a value of $0.001$ 
or to apply a constant kinematic viscosity on the order of 
$10^{-5}\,r^{2}_{0}\Omega_{0}$ (therefore altering the disk structure). 
Under these conditions, convergent migration resumes, but the high density
and strong torques prevent the planets from being captured in low degree 
resonances (e.g.,   2:1 and  3:2), and instead the system ends up 
in stable high degree mean-motion resonances (e.g.,  5:4 or  6:5). 
This behavior is illustrated in Figure~\ref{highdens}, where four 
cases with different initial density and viscosity parameters lead 
to trapping in high degree resonances. 
It is noteworthy that in the case with $\alpha = 0.001$ and 
$\Sigma_0 = 400\,\mathrm{g/cm^2}$ the planets are temporarily captured 
in  4:3 resonance; however, it is  broken on a short timescale by 
the strong torques exerted by the gas. The eccentricity evolution of 
the two planets is shown in Figure~\ref{ecce54} where at each 
resonance crossing a sudden jump in eccentricity is observed for 
both planets, as also discussed for some previous cases. 
When trapped in the 5:4 MMR, after an initial rapid growth, the orbital
eccentricity begins to decrease to low values, as in Figure~\ref{2au32res}.
This may  possibly be due to the inner planet approaching a steep density
gradient of the disk's gas while moving closer to the star. 
The damping of eccentricity suggests that this resonance may be stable 
over long timescales, even in the presence of the dissipative force due to 
the disk torques. 
A similar behavior is observed for the 6:5 resonance, and we tested with 
N-body calculations (aimed at mimicking the behavior after disk dispersal) 
that both resonances are stable at least over a timescale of $1\,\mathrm{Gyr}$.
We note that, in order to achieve capture in  6:5 resonance, we initialized 
the planets on more compact orbits compared to the other cases, just outside 
the 4:3 MMR. 
In this way the 5:4 MMR is approached when the planets are farther from 
the sun and the balance between the resonance strength and the tidal force
inducing migration is tipped in favor of the disk torque (compared to the other
cases), allowing the exterior planet to cross this resonance and attain 
the 6:5 MMR.

The simulations performed with \mouno\ and \modue\ show that resonant 
capture in low degree resonances occurs preferentially when the pair reaches
the resonance in the inner disk within a few au of the star, even for 
high values of  gas density, because  only then can the resonance 
forcing typically overcome the tidal forcing exerted by the gas. 
High-degree MMRs (i.e., compact orbital configurations above the 3:2 MMR)
are instead easier to attain when the pair of planets crosses the resonances 
in the outer disk, beyond several au from the star where tidal forces can 
more easily overcome the resonant forcing. Since the orbital migration of 
the pair after capture generally leads them inward, compact MMRs can also
be observed close to the star.
Therefore, the degree of  resonant configuration observed at present 
may provide some indication of the disk location at which capture occurred, 
and of the extent of the coupled inward migration. For example, a compact 
orbital configuration (e.g., 4:3 or 5:4 MMR) observed close to the star, 
around or inside $1$~au, may indicate that it was established farther out 
in the disk, and that the pair underwent a long-range orbital migration.


\section{Dust evolution near the resonant planets}
\label{sec:dust}

To model the dynamical features that two planets locked in resonance 
produce in the dust distribution around them, we integrated the trajectories 
of $400000$ dust grains. We started the integration when the planets begin 
to migrate, and considered both \mouno\ and \modue. We integrated the trajectories
of particles four different sizes: $10\,\mu\mathrm{m}$, $100\,\mu\mathrm{m}$, 
$0.1\,\mathrm{cm}$, and $1\,\mathrm{cm}$.
The Stokes numbers of the particles in these models, evaluated at about 
$5\,\mathrm{au}$ from the star, are $\lesssim 3 \times 10^{-3}$ when 
the gas surface density of the disk at $5\,\mathrm{au}$ is 
$\Sigma_{0}=1000\,\mathrm{g\,cm^{-2}}$. 
In the models with the lowest density, $\Sigma_{0}=50\,\mathrm{g\,cm^{-2}}$, 
the Stokes numbers have values  $\lesssim 7 \times 10^{-2}$. 


\begin{figure}
\includegraphics[width=\columnwidth]{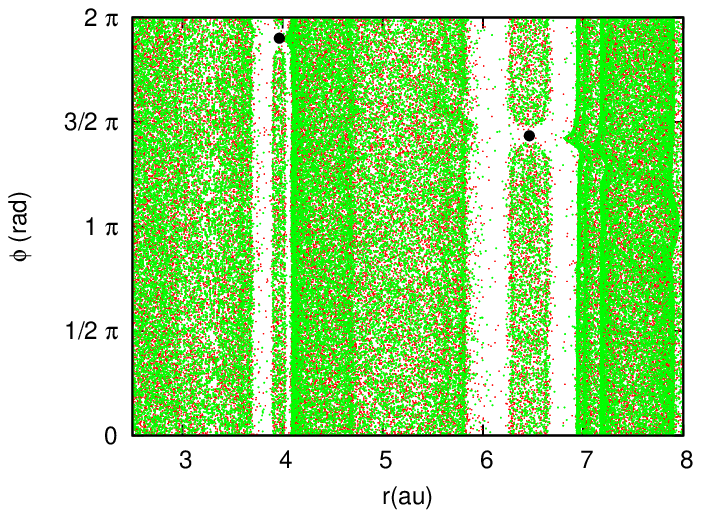}
\includegraphics[width=\columnwidth]{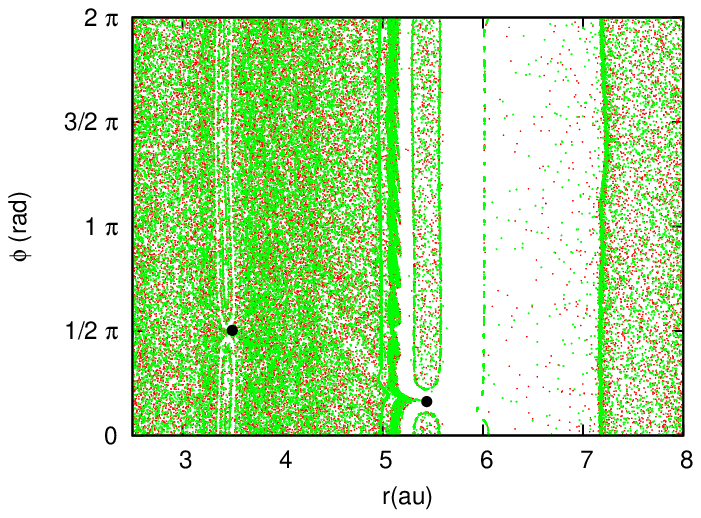}
\caption{\label{dust_21} 
Dust distribution around the planets trapped in  2:1 resonance 
at $t= 10\,\mathrm{Kyr}$ (top panel) and $t= 70\,\mathrm{Kyr}$ 
(bottom panel) in \modue. 
The black circles give the planets' positions.  
Only dust particles $10\,\mu\mathrm{m}$ and $1\,\mathrm{cm}$ in size 
are shown to avoid overcrowding the plot.
}
\end{figure}

We first consider the case when the pair is locked in  2:1 resonance in 
\modue, because the greater distance from the star allows  a better 
characterization of the dust features. 
In Figure~\ref{dust_21}, we show the distribution of $10\,\mu\mathrm{m}$ and
$1\,\mathrm{cm}$ dust particles in the case with 
$\Sigma_0 = 50\,\mathrm{g\,cm^{-2}}$ (\modue) after $10\,\mathrm{Kyr}$ from 
the beginning of the planet migration (top panel) and after $70\,\mathrm{Kyr}$
(bottom panel). Prior to resonance capture, two independent gaps in the dust
distribution start to develop around the two planets, where the distribution
appears slight denser at the outer edge of each gap (compared to the
density at their inner edge, see top panel of Figure~\ref{dust_21}). 
When the resonance is established, the migration rate of the inner planet
increases due to the tidal forcing exerted on the exterior planet 
(see discussion in Sect.~\ref{sec:model1}), and its gap slowly vanishes 
(see Figure~\ref{dust_21}, bottom panel). 

Meanwhile, the gap around the outer planet is not replenished and the region
behind the planet becomes depleted of dust. This effect may be
associated with the faster inward velocity of the planet compared to that of
the particles caused by gas drag.
Observations of wide gaps in the dusty disks, like the one in the bottom
panel of Figure~\ref{dust_21}, would be difficult to interpret since they
may be produced by a migrating low-mass planet or by a non-migrating 
massive planet.
A significant enhancement in the dust surface density is also observed at 
the inner border of the outer planet's gap, where dust is pushed inward. 
This dust feature is caused by the radial gradient of the perturbed pressure
that prevents dust from approaching the planet (by altering the local
rotation rate of the gas). The fact that solids can be collected ahead of
the planet as it migrates inward is another indication that 
the drag-induced drift of the dust is slower than the migration velocity
of the planet.

\begin{figure}
\includegraphics[width=\columnwidth]{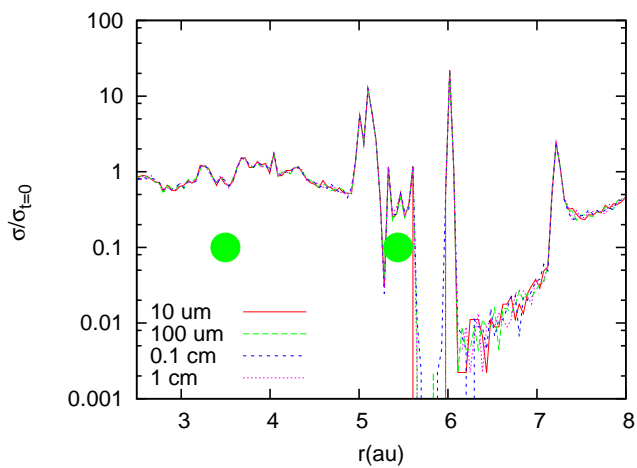}
\caption{\label{histo_2_21} 
Histogram of the dust surface density, normalized to that at $t=0$,
around a pair of planets trapped in  2:1 resonance
The plot refers to the same case as in Figure~\ref{dust_21}, at  time
$t= 70\,\mathrm{Kyr}$ (see bottom panel of that Figure). 
The green filled circles give the positions of the planets.  
The different line colors correspond to different grain sizes,
from $10\,\mu\mathrm{m}$ to $1\,\mathrm{cm}$. 
Only minor differences are observed among the various sizes
because all grains are well coupled to the gas.  
}
\end{figure}
To examine in some more detail the density distribution of the dust, 
we divided the radial range into a series of discrete radial bins 
so that, by counting the number of particles in each bin, the evolution 
of the surface density of dust ($\sigma$) can be monitored in an average
sense. 
In Figure~\ref{histo_2_21}, the value of $\sigma$ is shown after 
$t=70\,\mathrm{Kyr}$ in each radial bin, normalized to the corresponding
value at $t=0$. 
The main gap, extending beyond the orbit of the outer planet, is
approximately $1.5\,\mathrm{au}$ wide and its depth is on average 
about $1$\% of the initial dust density. 
The highest density feature, a dust ring around $6.1\,\mathrm{au}$, peaks at a value
more than ten times higher than the local initial density of the dust. 
This feature is maintained by the 6:5 mean motion resonance 
with the outer planet, following the planet during its inward
migration, and is populated by solids that filter through
the planet's orbit from the inner disk regions.
Farther out,
the region of enhanced density,
located at about $7.2\,\mathrm{au}$, is instead a remnant of the
dust orbiting at the outer border of the initial gap produced by
the planet, where dust was collected during the early phases of
evolution of the system.
It does not follow the planet during its orbital evolution 
because the inward drift of the dust due to gas drag is 
slower than  the planet migration speed and it is temporary in nature since 
the density slowly declines with time (toward the average value of
the surrounding distribution).
The wider dust density peak at about $5\,\mathrm{au}$ is instead clearly due 
to the sweeping effect operated by gas drag, in response to the perturbed
radial gradient of the gas pressure, and the perturbation moves inward along
with the planet.
There are only minor differences among the distributions of the different
particle species (see Figure~\ref{histo_2_21}). This is likely because the dynamical
evolution is dominated by drag effects on solids well coupled to
the gas (Stokes numbers are generally $\ll 1$, as mentioned above) and 
by the gravitational perturbations of the planets. 
A similar outcome was already noted in the simulations of \cite{marzari2019}.

\begin{figure}
\includegraphics[width=\columnwidth]{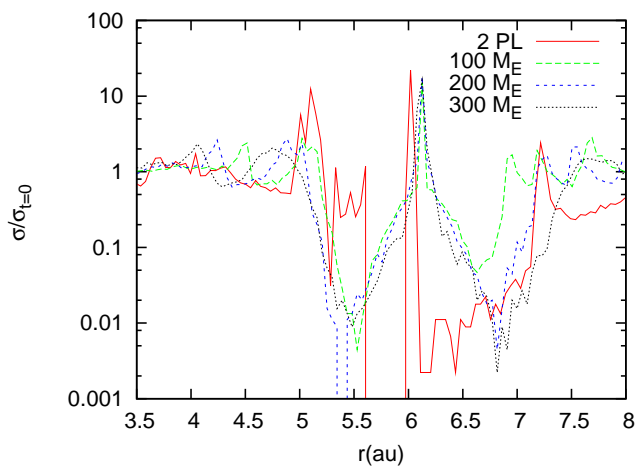}
\caption{\label{histo_gp} 
Comparison between the dust gap (normalized surface density) produced 
by two planets in  2:1 resonance for $10 \,\mu\mathrm{m}$ particles (see Figure~\ref{histo_2_21}) and 
those carved by a single massive planet. 
The red line shows the gap for the two-planet case, the green dashed line that
for a single planet of mass $100\,M_{\oplus}$, the blue dashed line that for 
a $200\,M_{\oplus}$ planet, and the black dotted line that for 
a $300\,M_{\oplus}$ planet.
}
\end{figure}

\begin{figure}
\includegraphics[width=0.68\columnwidth,angle=-90]{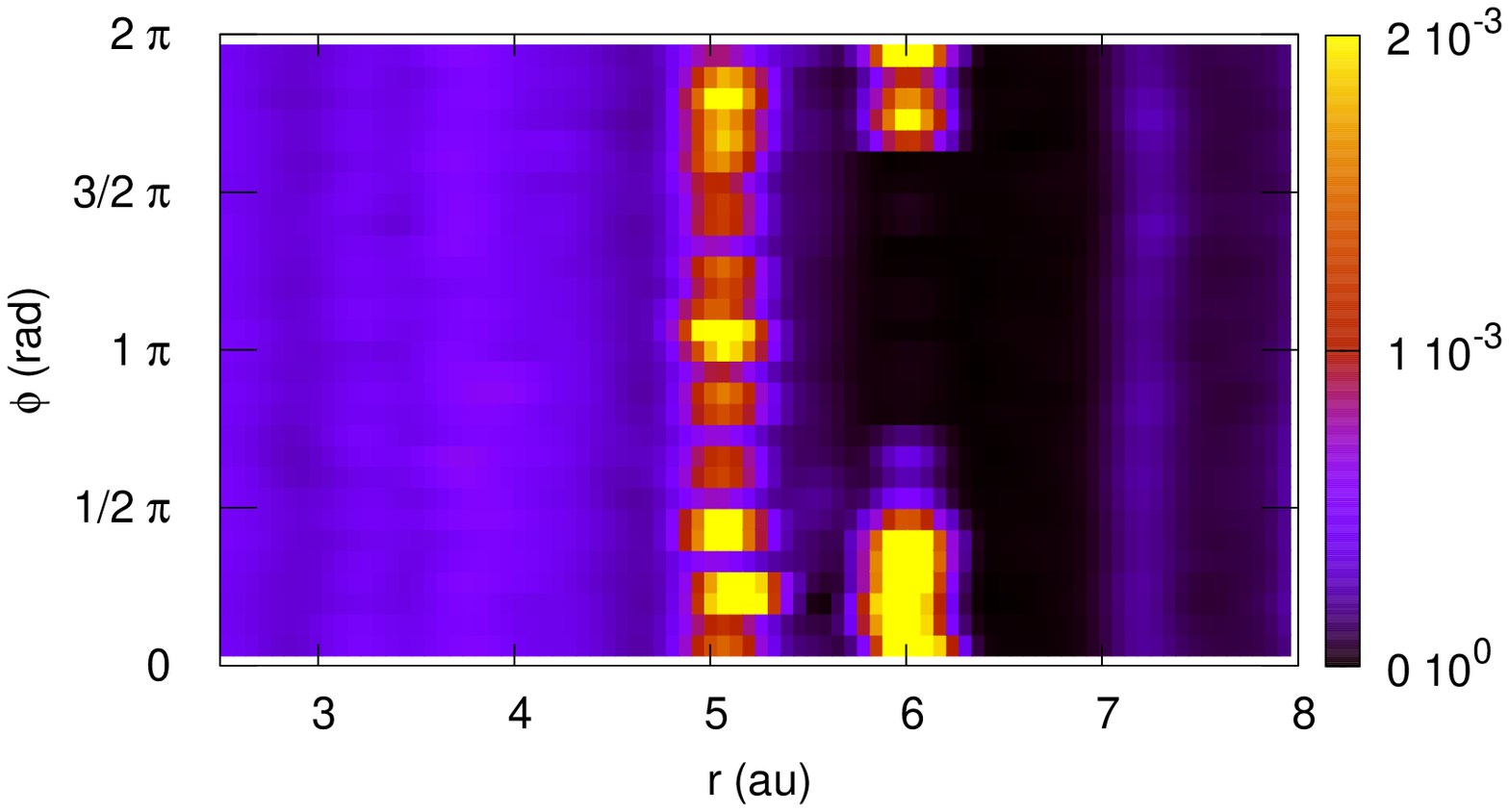}
\vskip -0.7 truecm
\includegraphics[width=0.68\columnwidth,angle=-90]{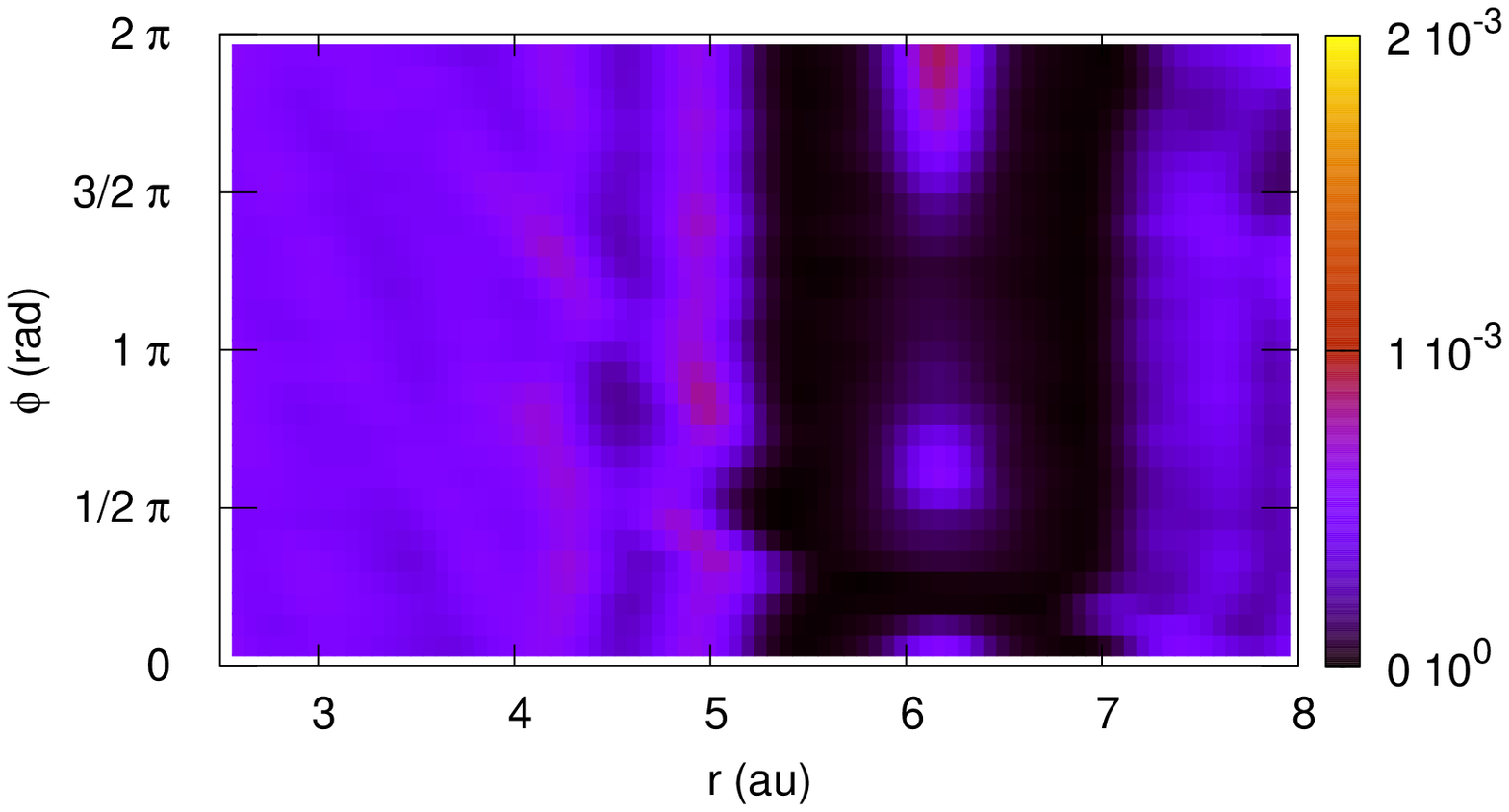}
\caption{\label{map} 
Smoothed maps of the dust distribution in the proximity of 
the planets. The top panel shows the density distribution,
normalized to the total number of dust particles in the model,
around two planets  in  2:1 resonance (Figure~\ref{dust_21} 
and Figure~\ref{histo_2_21}). 
The color bar range is in units of $10^{-3}$; the azimuthal
resolution is $0.2$~rad while the radial resolution is 
$0.15$~au. 
The bottom panel illustrates the dust distribution for 
a single planet with mass $m=200\,M_{\oplus}$.
As smoothing function, we applied a simple bell-shaped function \citep{lucy}.}
\end{figure}

From observations, the wide gap that develops around the pair 
of planets in resonance could be incorrectly interpreted as due 
to a single more massive planet. 
The similarity of the large-scale dust distributions for these
cases is illustrated in Figure~\ref{histo_gp}, where 
the normalized surface density of the dust due to two planets
orbiting in  2:1 resonance (those in Figure~\ref{histo_2_21})
is compared to that of a single planet with mass equal to $100$,
$200$, and $300\,M_{\oplus}$. 
The physical properties of the gas disk are the same in all
simulations. 
There are significant differences in the morphology of the gap
(and of other small-scale features) in the dust formed by the 
two planets compared to what a single and more massive planet
would produce.
However, given the scale of these features, they would 
be difficult to detect with current instruments. 
For the cases with a single planet, the outer edge of the dust 
gap becomes wider as the planet mass increases, a behavior reflecting
the formation of a wider gap in the gas as tidal perturbations
increase. 
Overall, the width of the dust gap produced by a single
$200\,M_{\oplus}$ planet is comparable to that generated by 
the two planets in resonance. 
The amount of dust depletion over the region is also comparable. 
This is further confirmed by the two smooth density maps shown 
in Figure~\ref{map}, where the case with two planets (in a 2:1
resonance) is compared to that with a single planet with 
$m = 200\,M_{\oplus}$. In the first case (top panel), 
the inner border of the gap is denser due to sweeping 
and trapping of dust during the inward migration of the outer
planet. In both cases, particles remain trapped in horseshoe
orbits, although the density of solids appears higher in 
the case with two planets. This is probably caused because the two smaller planets in resonance migrate a substantial
distance and can collect solids filtering from the interior 
toward the exterior of the orbits. The single more massive planet
clears a gap much more rapidly and is more effective at depleting
the horseshoe region of dust. This effect is somewhat artificial
since we do not model the growth of the planet which, 
if slow enough, would help increase the dust density in 
the region.
It is clear from the previous plots that when interpreting
large-scale features observed in dust distributions around stars, particular care is necessary before reaching conclusions based 
on perturbations effects in single-planet models.

\begin{figure}
\includegraphics[width=\columnwidth]{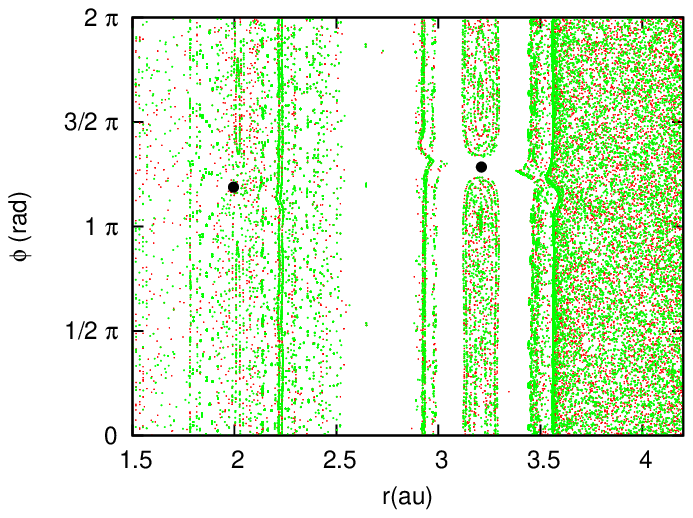}
\includegraphics[width=\columnwidth]{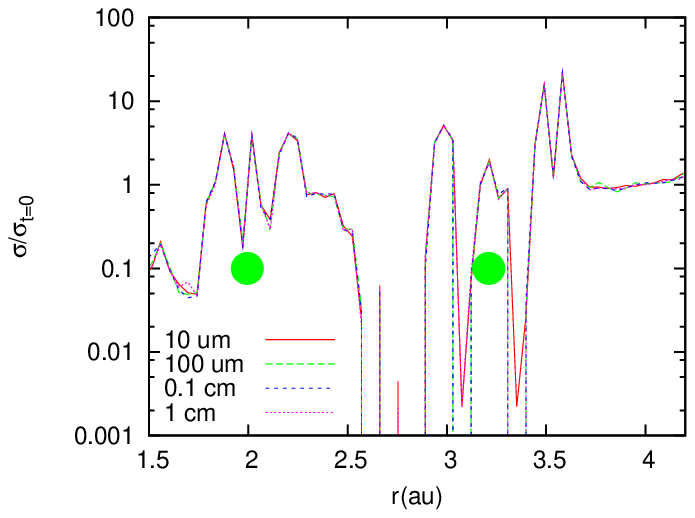}
\includegraphics[width=\columnwidth]{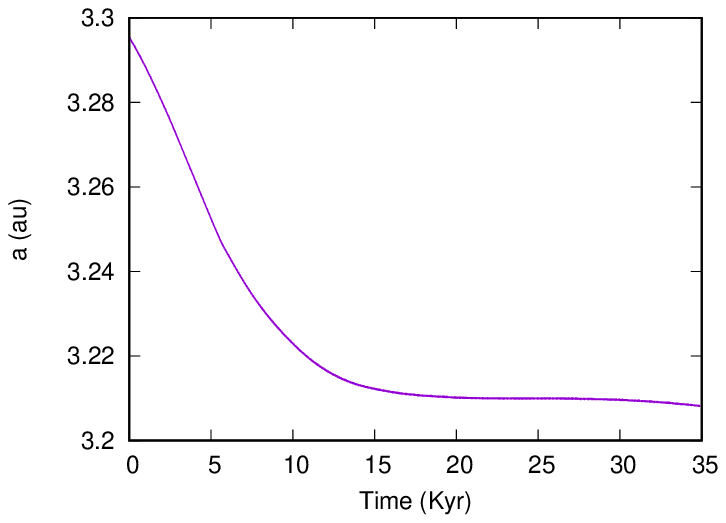}
\caption{\label{dust_21_mouno} 
Top panel: Dust distribution around planets locked in   2:1
resonance at $t=40\,\mathrm{Kyr}$ in a \mouno\ with
$\Sigma_0=50\,\mathrm{g\,cm^{-2}}$. The black circles give the planets' positions.
Middle panel: Histogram of the dust surface density where
the planets' orbital radius is given by a green filled circle.  
Bottom panel: Semi-major axis evolution with time of the outer planet   
showing a period of stalled migration when the pair of planets are trapped 
in resonance. 
}
\end{figure}
The behavior of the dust appears different, but with some similarities,   
in the case with low density ($\Sigma_0=50\,\mathrm{g\,cm^{-2}}$) in \mouno\
(see Figure~\ref{dust_21_mouno}). The two planets, once trapped in  2:1
resonance, experience an extended period of stalled migration during which their
semi-major axes are nearly stationary (see bottom panel of
Figure~\ref{dust_21_mouno}).
The exterior border of the outer planet's gas gap acts as a barrier to dust
particles migrating inward, and they accumulate at that location (see top and
middle panels of Figure~\ref{dust_21}). 
Between the two planets, between $2.5$ and $3\,\mathrm{au}$, the dust density 
is significantly reduced (to $\sim 0.1$\% of the initial value) likely due to 
the barrier effect operated by the outer planet, which prevents dust grains from
crossing inside  the planet's orbit. 

Concurrently, dust drifting inward is halted exterior to the inner planet's
orbit (again due to the perturbed pressure's radial gradient) where grains 
 collect and form a mild density excess, approximately a factor of five 
as large as the local initial density.
In this case too a wide gap may be detected by observations whose dynamical 
origin, however, is somewhat different from that illustrated in
Figure~\ref{dust_21} (exterior to the outer planet, see bottom panel).
In that case,  the migration speed of the planets is probably 
responsible for the widening of the gap (see also    Figure~\ref{dust_21_mouno},  where the migration of the pair is nearly stalled).
In the case shown in Figure~\ref{dust_21} 
only two high-density dust rings are produced; instead,    in the case shown in
Figure~\ref{dust_21_mouno} there are three: 
one trailing the inner planet and two around the outer planet.
In both configurations the density peaks trailing the outer planet may be 
the most significant features (from an observational standpoint) in 
the dust distributions, exceeding the surrounding density levels by factors 
$\gtrsim 10$ (see histograms in Figures~\ref{histo_2_21} and \ref{dust_21_mouno}).
In the other simulations based on \mouno, with capture in resonance, 
the planets rapidly move too close to the star because of the high values of
$\Sigma_{0}$. These configurations do not appear to show features amenable to
observations (over the timescales of our simulations).

\begin{figure}
\includegraphics[width=\columnwidth]{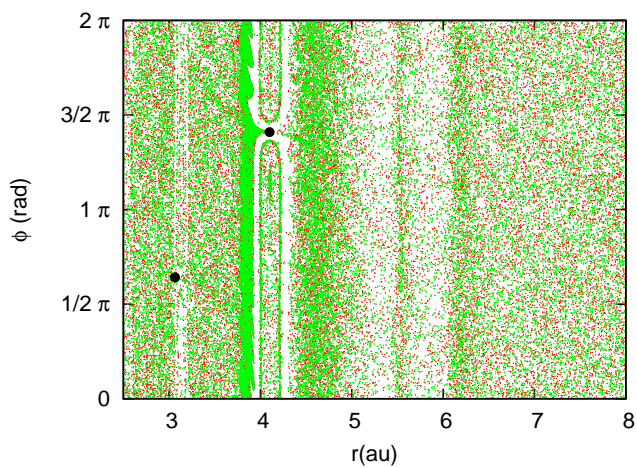}
\caption{\label{dust_32} 
Dust distribution at $t=100\,\mathrm{Kyr}$ for the two planets locked
in  3:2 resonance (\modue). 
}
\end{figure}

In the \modue\ case with capture in  3:2 resonance, the inward drift 
of the planets is faster compared to the case with capture in the 2:1 MMR 
since the disk density is higher and the features in the dust distribution are
more complex (see Figure~\ref{dust_32}). 
In this case the inner planet forms only a very shallow gap in the surrounding 
dust since the inward migration prevents the formation of large radial
gradients in gas distribution. Contrary to the case shown in Figure~\ref{dust_21},
the outer planet does not develop a deep and extended gap in the dust, but 
it does show significant density enhancements in the dust interior and exterior
of the planet's orbit: the density $\sigma$ at these peaks ranges from $10$ 
to $30$ times above the local initial values. 
Beyond the outer planet, at greater radial distances, there appear to be 
low-density regions, where the sigma value is about $20$\% less than the initial
values.
In this case, the dust signatures induced by the planets in resonance are 
less pronounced (with respect to the previous cases) and may be more difficult 
to detect.

\begin{figure}
\includegraphics[width=\columnwidth]{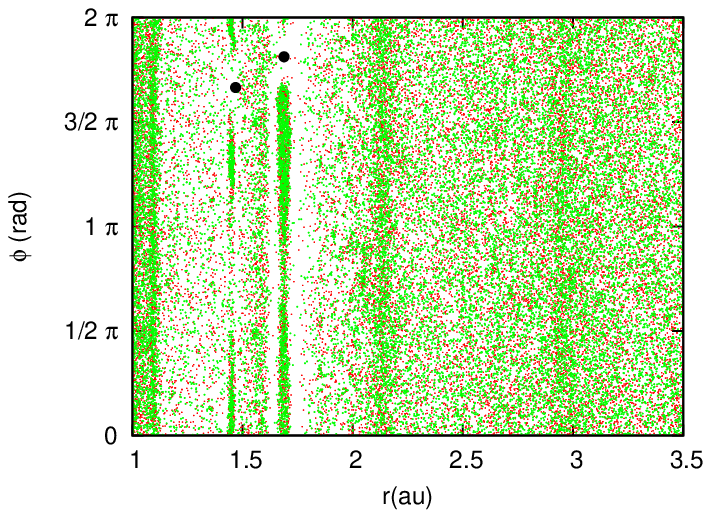}
\includegraphics[width=\columnwidth]{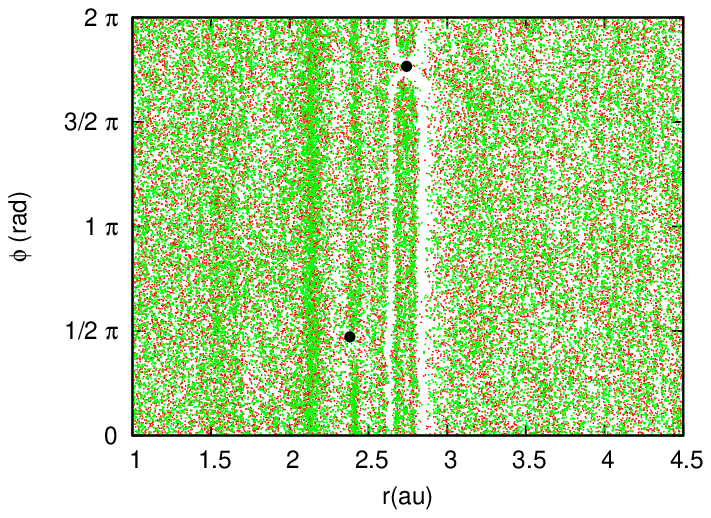}
\caption{\label{dust_45} 
Dust distributions around two planets trapped in  5:4 resonance 
($\Sigma_0 = 400\,\mathrm{g\,cm^{-2}}$, $\alpha=0.001$) at $t=60\,\mathrm{Kyr}$
(top panel) and around two planets trapped in  6:5 ($\Sigma_0 = 400\,\mathrm{g\,cm^{-2}}$, $\alpha=0.001$) at $t=48\,\mathrm{Kyr}$ 
(bottom panel). 
}
\end{figure}
When the planets are trapped in higher-degree resonances (e.g.,  5:4 and 6:5), 
the signatures in the surrounding dust distributions are relatively minor, 
as illustrated in Figure~\ref{dust_45}. 
In these models the absence of gaps in the dust is probably related to the
fast migration velocity because, in \modue, the capture in such compact orbital 
configurations occurs when the disk surface density is high 
($\Sigma_0 \geq 400\,\mathrm{g\,cm^{-2}}$). 
Consequently, the gravitational perturbations of the planets are less effective 
at perturbing the pressure gradient of the gas and the efficiency in collecting 
dust is significantly diminished.
In the case of the 5:4 MMR (top panel of Figure~\ref{dust_45}) a common gap is
formed, but its depth is only about a factor of $3$ smaller than the local
initial dust density, a difference that is likely difficult to  detect
by observations.

In summary, our simulations indicate that only wide resonant configurations, like 
the 2:1 MMR, may induce significant features in the surrounding dust distribution
that can be more likely probed by observations.
When the planets are locked in higher-degree resonances and migrate fast, 
they are more inefficient at perturbing the dust. 

An additional noteworthy feature of the dust distributions is 
the enhanced dust density at the borders of the gap carved by the planets 
locked in  2:1 and 3:2 resonances. At these locations the concentration 
of millimeter- to centimeter-sized solids may increase tenfold relative to the initial dust 
distribution, potentially triggering streaming instability events. In turn, these 
events may lead to the formation of a significant population of planetesimals. 
This phenomenon was recently modeled with numerical simulations by \cite{erik2020A&A...635A.110E}. 

\section{Summary and discussion}
\label{sec:discussion}

We performed hydrodynamic simulations of planet--disk interactions in multi-planet
systems. We considered a pair of super-Earths embedded in disks of various gas
densities and viscosities, where the exterior planet is the more massive. 
This configuration 
should favor convergent migration because the disk's thermo-dynamical 
state is such that a stronger tidal torque is exerted on the outer planet
than is on the inner one. 
By adding dust particles to the simulations, we also computed their trajectories 
in the proximity of the planets in order to test the influence of the resonant
configurations on the dust distributions. 
We considered two different setups, where we changed the initial distance of 
the planets from the star. In \mouno, the inner planet is initially located at 
$2\,\mathrm{au,}$ while in \modue\ the inner planet starts from $4\,\mathrm{au}$. 
These two models provide a different balance between the resonance strength 
and the tidal forcing (by the gas) at the locations of resonance capture.

We find that resonance trapping always occurs in first-order resonances, even 
of high degree, for different values of the initial gas density and viscosity.
However, convergent migration may be inhibited either by a temporary outward
migration of the outer planet or by a slowdown of its inward migration
compared to that of the inner planet. We did not investigate these configurations
in detail, but it appears that either of these two outcomes may be due to 
mutual perturbations between the planets. In \modue, divergent migration tends
to occur for high values of the gas density and of the $\alpha$-viscosity 
parameter. 
A reduction of the viscosity parameter or the use of a constant kinematic 
viscosity $\nu$ allows convergent migration to resume.

Close to the star, the capture in low degree resonances (e.g.,   2:1 or  3:2) 
is favored because resonant interactions scale as $\Omega^{2}$, whereas the tidal
forcing in these calculations is $\propto r^{1/2}$. In \modue, where the distance
from the star is greater, trapping in high degree resonances (5:4 or 6:5)
can occur. 
N-body calculations indicate that these compact resonances are dynamically stable, 
once the gas dissipates, over timescales on the order of several billion years. 
This behavior suggests that
low degree MMRs presently detected among super-Earths may have occurred when the
planets were on converging orbits and became trapped relatively close to the star. 
Therefore,
their formation sites might have been within a few to several au from the star. 
Instead, high degree resonances are generally expected to occur far from the star,
at several au, suggesting formation at larger radii. The degree of the resonance 
may thus provide some indication about the formation distance from the star of 
the pair.

In the range of planetary masses adopted in our models, a gap in the dust is 
expected to form on a short timescale if a single non-migrating planet is considered. 
The situation may be different when considering a system of two planets that  are allowed to migrate. If the orbits of the pair converge, they can be trapped in resonance and 
the surrounding dust distribution can be significantly affected. 
Once in resonance, the inner planet may be forced to migrate at a faster rate (than
it does prior to capture in the MMR), due to the gravitational push by 
the outer more massive planet. Consequently, the dust gap around the inner planet may 
be progressively erased. This outcome is shown in simulations where the planets 
become trapped in the 2:1 MMR. In this scenario, the presence of the inner planet
cannot be detected by looking for dust gaps through observations of disks. 
At larger radial distances, the gap around the exterior and more massive planet steadily
widens because of two distinct and mutually exclusive mechanisms. In the first case (see
Figure~\ref{dust_21}), the pair of planets migrate quickly and the dust does not drift
inward fast enough to replace the dust swept ahead of the planet, which leads to 
the formation of a wide gap extending beyond the exterior planet. 
In the second case (see Figure~\ref{dust_21_mouno}), once trapped in resonance
the planets undergo a period of stalled migration and the outer planet acts as a barrier to 
the dust grains drifting inward. In this latter case, a large gap in the dust develops 
 between the two planets since the local dust has drifted toward the inner planet.  
In both cases the large gap developing around the planets' orbits might be mistaken for 
that produced by a single giant planet.

Higher-degree  resonances appear to have less impact on the dust distribution. 
The 3:2 resonance leads to the formation of a region of enhanced density of dust grains
in front of the outer planet and of a depleted region behind it, likely a remnant of the
early migration phases of the pair. The 5:4 and 6:5 resonances are even less effective in sculpting 
the dust distribution, possibly because to achieve these resonant configurations 
the planets need to migrate inward quite rapidly. In the long term, if the migration 
of the pair locked in resonance reduces or stalls, it is likely that the planets will open 
a single wide gap in the dust distribution because of the close proximity of 
their orbits.

Due to the large computational overhead of hydrodynamic simulations, we only sampled
a limited number of disk parameters, planetary masses, and initial orbits. 
Different density slopes $d\ln{\Sigma}/d\ln{r}$ of the gas, different values of kinematic 
viscosity $\nu$, and disk thermal states may change the migration pattern of the planets.
The occurrence of convergent or divergent migration is significantly affected by these
parameters, and the convergent migration may occur at different locations in the disk
depending on the parameter choice. 
The migration speed is another important aspect influencing the type of resonance that 
is established and the dynamical response of the dust distribution around the pair 
of planets. 
However, the partial disappearance of the gap around the inner planet, the formation 
of a wide gap around the outer planet for the 2:1 MMR, and the less conspicuous features 
that appear in the dust for higher-degree resonances are robust results.

\begin{acknowledgements}
We would like to thank the referee, Alexander Mustill, for his comments
and suggestions.
G.D. acknowledges support from NASA’s Research Opportunities in Space 
and Earth Science (ROSES) through NASA's Exoplanets Research Program 
(proposal 80HQTR19T0086).
Computational resources supporting this work were provided by the NASA
High-End Computing (HEC) Program through the NASA Advanced Supercomputing
(NAS) Division at Ames Research Center.
\end{acknowledgements}

\bibliographystyle{aa}
\bibliography{biblio}


\appendix
\section{Effects of stellar irradiation}
\label{sec:irr}

\begin{figure*}
\resizebox{\linewidth}{!}{\includegraphics[]{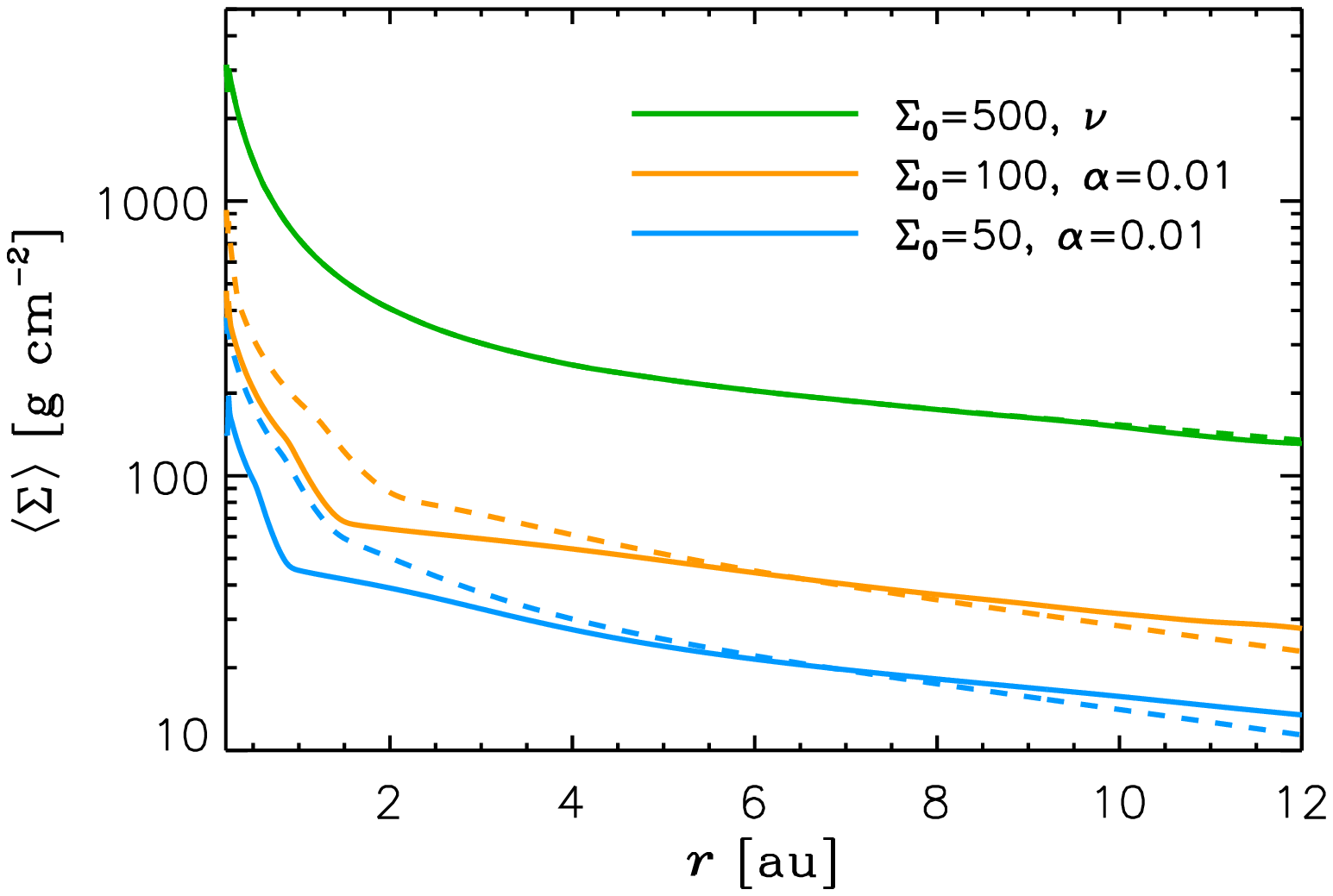}%
                          \includegraphics[]{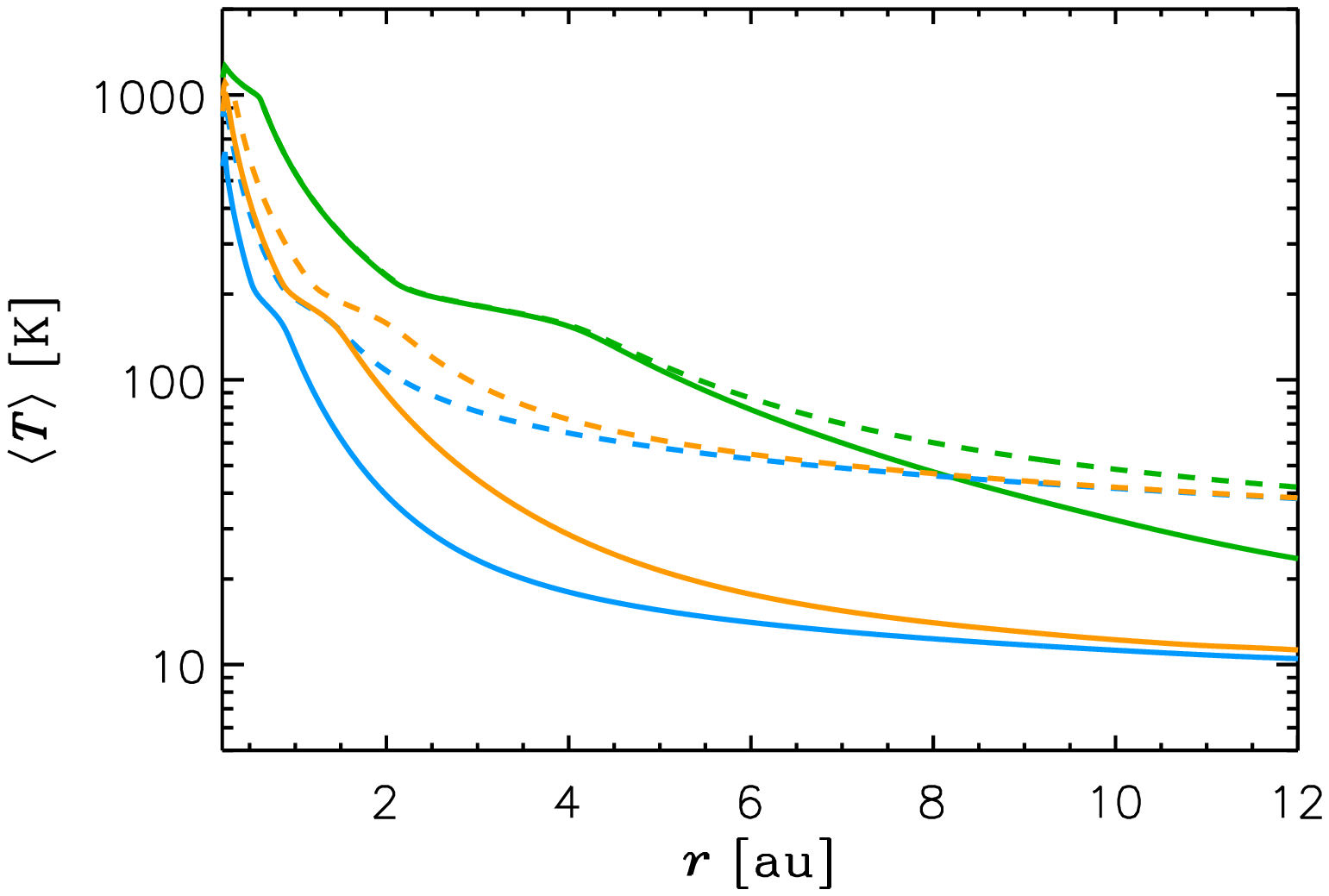}}
\caption{\label{fig:nop_si}
Average gas density (left) and temperature (right) of the unperturbed 
disks after $12$~$\mathrm{kyr}$. The solid curves represent the models 
shown in Figure~\ref{fig:nop} (see legend); the  dashed curves refer 
to the corresponding models that include stellar irradiation, as outlined 
in the text.
As expected, beyond some distance from the star, gas temperature is 
determined by the stellar irradiation temperature.
}
\end{figure*}
Irradiation from the central star affects the thermal structure of a
circumstellar disk and essentially determines its temperature beyond some
distance from the star.
The lower the disk's density $\Sigma$ and the smaller the kinematic viscosity 
$\nu$, the shorter  the radial distance at which heating by stellar radiation
can significantly impact the thermal balance of the gas. 
In order to asses possible effects of irradiation in the disk models 
discussed above, the right-hand side of Equation~(\ref{eq:eneq}) is modified
to include a stellar heating term $Q_{\mathrm{irr}}$, as in Equation~(10) 
of \citet[][where the term $Q^{-}$ is also modified accordingly; see discussion]{dangelo2012}.

In the 2D $(r,\phi)$ geometry, irradiation cannot be accurately taken into
account due to the lack of the disk's vertical structure, which is needed to
estimate energy deposition in a vertical column of gas at a given radial 
distance \citep[see, e.g.,][and references therein]{bitsch2013}. 
Our implementation 
is therefore approximate and follows an approach often applied in prior
studies \citep[e.g.,][and references therein]{dubus1999,menou2004}, with 
some simplifications outlined in \citet{hueso2005}.
We assume that the stellar effective temperature and radius are equal to
$4300\,\mathrm{K}$ and $2.6\,R_{\sun}$, respectively.

Stellar irradiation is expected to affect a wider range of distances,
extending inward, as gas density and kinematic viscosity decrease.
Figure~\ref{fig:nop_si} includes some results of Figure~\ref{fig:nop} 
(solid lines, see legend) and equivalent models that account for 
irradiation (dashed lines).
As anticipated above, gas temperature is determined by the stellar 
irradiation temperature beyond a radial distance where viscous heating
$Q^{+}\ll Q_{\mathrm{irr}}$.
The disk structure in
our models with $\Sigma_{0}\gtrsim 500\,\mathrm{g\,cm^{-2}}$ is not 
significantly affected by stellar irradiation over the range of radial
distances considered here.

\begin{figure}
\resizebox{\linewidth}{!}{\includegraphics[]{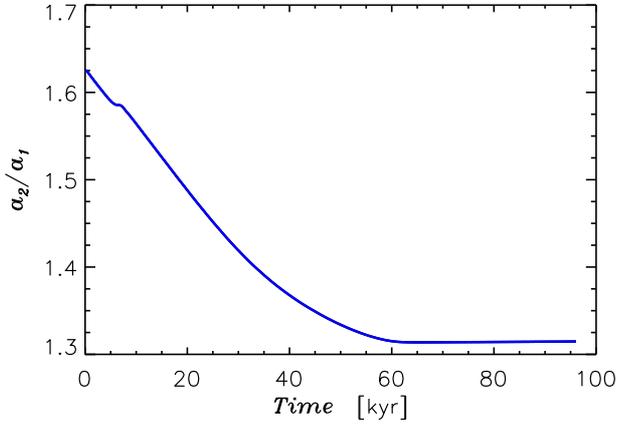}}
\caption{\label{fig:arat100_irr}
Ratio of the semi-major axes of a pair of planets using the disk configuration
with $\Sigma_{0}=100\,\mathrm{g\,cm^{-2}}$, turbulence viscosity
corresponding to $\alpha=0.01$, and including stellar irradiation. 
The planets are initiated according to the setup in \modue.
The outer planet transits the 2:1 MMR with the inner planet and becomes 
trapped in the 3:2 MMR, as in the model that does not account for irradiation
(see Sect.~\ref{sec:model2}).
}
\end{figure}
Referring to Figure~\ref{fig:nop_si}, in the highest density model 
(green curves) $\Sigma$ is essentially unaffected by irradiation (out
to $12$~au) and $T$ starts deviating from the no-irradiation model at
$r>6$~au. 
Since, in \modue, the outer planet starts its evolution 
at $6.5$~au, differences in the orbital evolution of the pair might 
not be large.
The situation is expected to be
different for the lower density configurations, since the difference in
temperature is significant even at short orbital distances from the star.
The higher temperature produces a higher aspect ratio $h$, which would
contribute to a slower migration velocity (possibly of both planets).
This reduction may be partly offset by the somewhat larger $\Sigma$
at  small $r$ (compare orange and blue solid and dashed curves in the left panel
of Figure~\ref{fig:nop_si}).
However, the configuration with $\Sigma_{0}=50\,\mathrm{g\,cm^{-2}}$ 
results in capture in the 2:1 MMR in the model without irradiation and is
therefore expected to produce the same resonant orbits in the model with
irradiation.
The configuration with $\Sigma_{0}=100\,\mathrm{g\,cm^{-2}}$, which 
produces a 3:2 MMR in the no-irradiation model, may in principle result 
in a less compact MMR in the model with irradiation.
We performed a calculation of this particular configuration and show 
the results in Figure~\ref{fig:arat100_irr}. 
Despite the changes in the radial distribution of temperature and of
$\Sigma$, the evolution of the pair results again in the
capture in the 3:2 MMR. The plot indicates that the exterior planet is 
briefly caught in the 2:1 MMR with the interior planet, before skipping 
the resonance and continuing to migrate inward. 
During the temporary capture in the 2:1 MMR, the orbital eccentricity of
the two planets remains low ($< 0.01$). Since resonance breaking  is
not aided by an increase in eccentricity, the temporary capture can be
considered stochastic in nature.

Clearly, there may be additional and more subtle effects brought about 
by the warmer gas, although in general it is expected that disk regions 
farther from the star are affected more than are the inner disk regions 
(e.g., as in our \mouno). 
We plan to investigate  in greater detail the effects of irradiation in
low-density low-viscosity disks, including resonance capture of a pair 
of planets and dust evolution, in subsequent work.

\end{document}